\documentclass[aip,jcp,
 reprint,
superscriptaddress,
amsmath,amssymb,
]{revtex4-2}

\usepackage{graphicx}
\usepackage{dcolumn}
\usepackage{bm}
\usepackage{xfrac}
\usepackage{color}

\begin{document}


\title{A patchy-particle 3-dimensional octagonal quasicrystal}

\author{Akie Kowaguchi}
\affiliation{Physical and Theoretical Chemistry Laboratory, Department of Chemistry, University of Oxford, South Parks Road, Oxford, OX1 3QZ, United Kingdom}
\affiliation{Department of Mechanical Engineering, Keio University, Yokohama 223-8522, Japan}

\author{Savan Mehta}
\author{Jonathan P. K. Doye}
\email{jonathan.doye@chem.ox.ac.uk}
\affiliation{Physical and Theoretical Chemistry Laboratory, Department of Chemistry, University of Oxford, South Parks Road, Oxford, OX1 3QZ, United Kingdom}

\author{Eva G. Noya}
\affiliation{Instituto de Qu\'{i}mica F\'{i}sica Blas Cabrera, Consejo Superior de Investigaciones Cient\'{i}ficas, CSIC, Calle Serrano 119, 28006 Madrid, Spain}

\date{\today}

\begin{abstract}
We devise an ideal 3-dimensional octagonal quasicrystal that is based upon the 2-dimensional Ammann-Beenker tiling  and that is potentially suitable for realization with patchy particles. Based on an analysis of its local environments we design a binary system of 5- and 8-patch particles that in simulations assembles into a 3-dimensional octagonal quasicrystal. The local structure is subtly different from the original ideal quasicrystal possessing a narrower coordination-number distribution; in fact, the 8-patch particles are not needed and  a one-component system of the 5-patch particles assembles into an essentially identical octagonal quasicrystal.  We also consider a one-component system of the 8-patch particles; this assembles into a cluster with a number of crystalline domains, but which, because of the coherent boundaries between the crystallites, has approximate eight-fold order.
We envisage that these systems could be realized using DNA origami or protein design.
\end{abstract}

\maketitle

\section{Introduction}
\label{sect:Intro}

Quasicrystals (QCs) have long-range order, as exemplified by the sharp peaks in their diffraction patterns, but without a periodically repeating unit cell. QCs can thus have symmetries that are not possible for periodic crystals. The first example was discovered by Shechtman for an Al/Mn alloy and had icosahedral symmetry.\cite{Shechtman84} This discovery stimulated a search for further examples and quasicrystalline alloys with decagonal,\cite{Bendersky85} dodecagonal\cite{Ishimasa85,Iwami15} and octagonal\cite{Wang87,Cao88} symmetry were quickly discovered. More recently, QCs have also been experimentally discovered in soft matter systems,\cite{Dotera11,Zhou23,Wang23b} but these have been generally limited to dodecagonal symmetry. 
Thus, discovering new ways to form QCs and increasing the possible repertoires of structures and symmetries is of significant interest.

Theory and simulations have the potential to significantly contribute to this goal by identifying how interparticle interactions can be designed to achieve quasicrystallinity. QCs  involve multiple length scales that are related by a specific irrational ratio, e.g.\ the golden ratio 
in the case of icosahedral QCs. This feature is exploited in approaches that use isotropic potentials with multiple length scales.\cite{Dotera14,Savitz18} For example, a theoretical framework for designing such ultra-soft potentials has been developed.\cite{Subramanian16,Ratliff19,Archer22} 
QCs have also been achieved for multiple length-scale potentials with hard cores, albeit more through the systematic exploration of the space of potential parameters, both in 2D\cite{Engel07} and 3D.\cite{Engel15,Damasceno17,Dshemuchadse21} 
Inverse design methods also hold significant promise for quasicrystal discovery.\cite{BedollaMontiel24}

An alternative approach to design QCs is to attempt to directly design the tendency to form a particular symmetry through directional bonding.\cite{vanderLinden12,Reinhardt16,Tracey21,Noya21,Noya25,Pinto25}  
The idea is that the directionality helps to induce both local and global non-crystallographic symmetry, where the quasiperiodicity is then a consequence of the formation of that global symmetry.
This approach has recently been successfully applied to design patchy particles that form icosahedral quasicrystals in simulations.\cite{Noya21,Noya25} 
First, it is important that the patch geometry is designed to be consistent with the desired point group symmetry; e.g.\ for the icosahedral QC-forming systems the patches are directed along subsets of the symmetry axes of $I_h$. Although this encourages motifs with the desired local orientational order, it is also important that the system can form a globally ordered structure that is sufficiently stable to out-compete possible periodic crystal forms during assembly. In the icosahedral QC-forming system, this second feature was achieved by basing the patch design on the local environments in an ideal target QC (that was obtained by the cut-and-project approach), thus helping to ensure that the best way that the local environments determined by the patch geometry combine together is as a QC with the target global symmetry.

Octagonal symmetry is one of the more experimentally rarely observed quasicrystalline symmetries with only a few examples in alloys.\cite{Wang87,Cao88} Moreover, in these examples, the quasicrystalline phase is metastable, being observed to transform to a $\beta$-Mn-type crystal.\cite{Wang90} The number of examples of octagonal quasicrystal self-assembly in simulations is also modest, all in systems with isotropic interactions, both in 2D\cite{Zu17,Fayen23} and in 3D.\cite{Damasceno17}
Here, the goal is to increase the potential means to realize octagonal QCs by designing a system that can form a 3-dimensional octagonal QC through directional bonding.

\begin{figure*}[t]
\includegraphics[width=18.0cm]{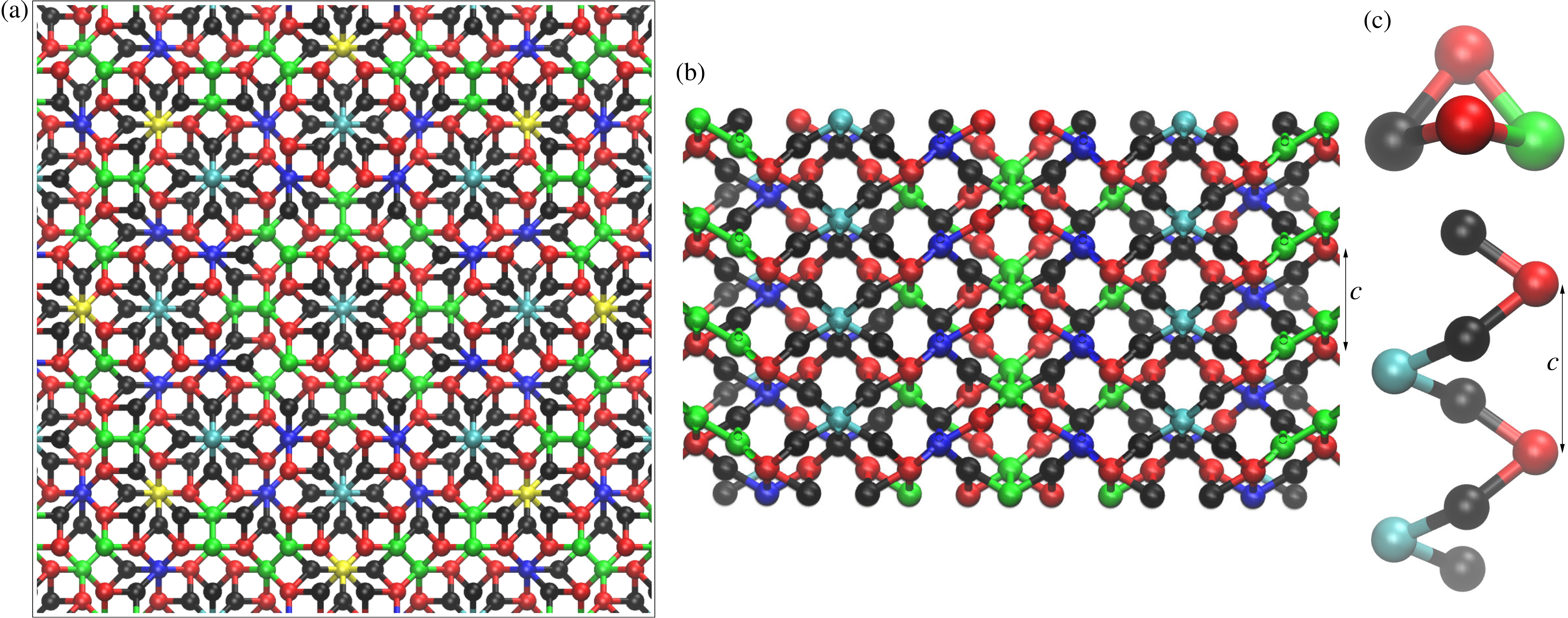}
	\caption{\label{fig:ideal} The ideal target octagonal quasicrystal. (a) The structure projected down the $c$-axis is that of an Ammann-Beenker tiling with particles at the corners of the square and rhomboidal tiles. The particles are coloured by their coordination number (See Fig.\ \ref{fig:ideal_env}). 
(b) A side-view along one of the two-fold axes.
(c) In 3-dimensions, the square is puckered and the rhombus corresponds to a right- (illustrated) or left-handed helix with a pitch length of the $c$ repeat. 
}
\end{figure*}

\begin{figure}[t]
\includegraphics[width=8.4cm]{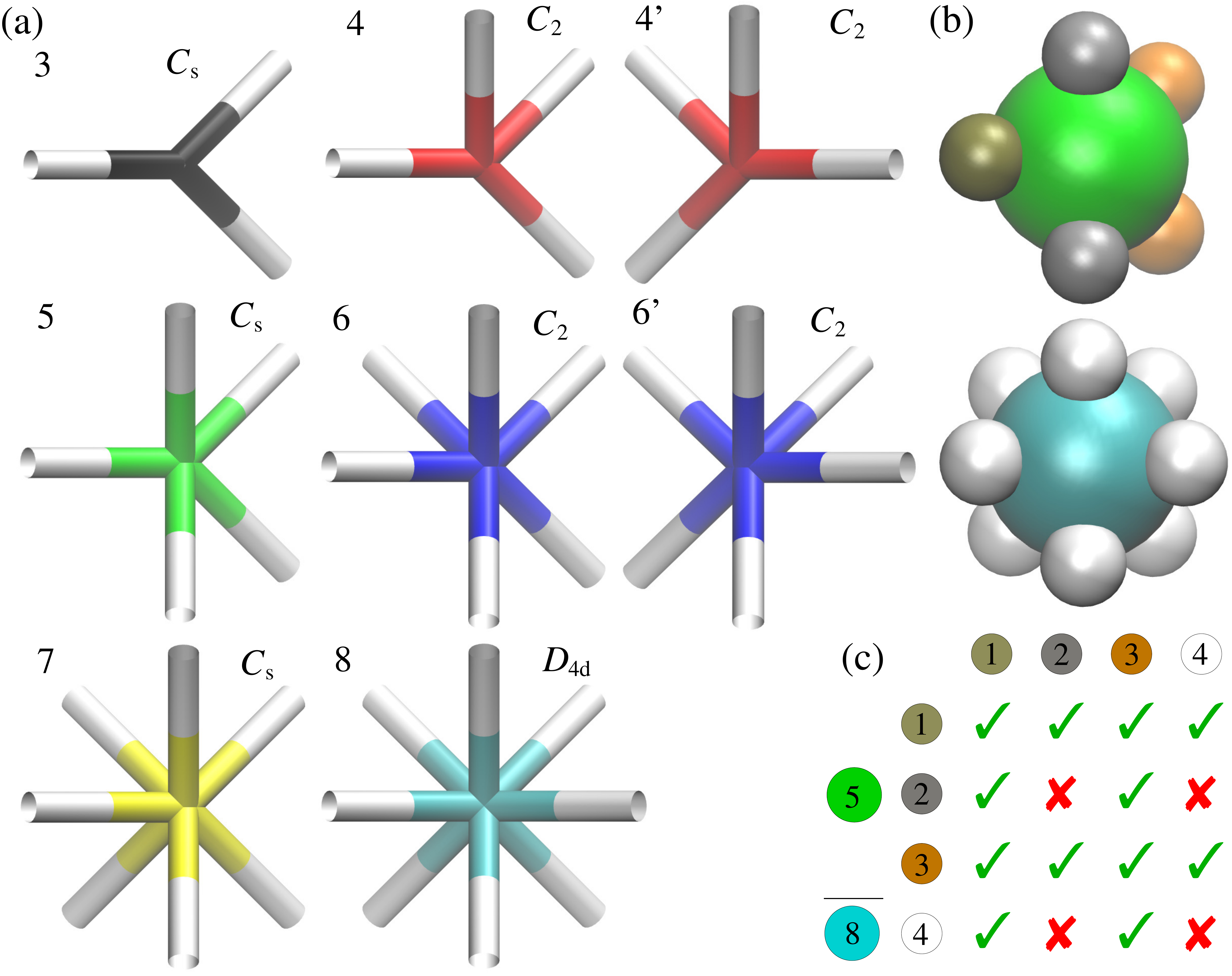}
\caption{\label{fig:ideal_env} 
(a) The coordination environments in the ideal quasicrystal labelled by their coordination number and their local point group symmetry. The two four-coordinate and six-coordinate environments are enantiomeric. 
(b) The two patchy particles.
(c) The interaction matrix between the four different types of patches on the above two particles. 
}
\end{figure}

\section{An ideal octagonal quasicrystal}
\label{sect:ideal_octQC}
To apply this approach, we first need an ideal 3D octagonal QC to provide a basis for the patchy-particle designs. 
Our starting point is the 2D Ammann-Beenker tiling \cite{Grunbaum} that can, for example, be derived by projection of a 4D hypercubic lattice into 2D (Supplementary Section S1). 
The tiling consists of squares and rhombi (with internal angles of 45$^\circ$ and 135$^\circ$) with the edges of these polygons being equally likely to be oriented along eight equivalent directions. The simplest way to create a 3D octagonal QC from this tiling would be to place particles at the vertices and have a simple periodic stacking of identical Ammann-Beenker layers
(as was done in Ref.\ \onlinecite{Tracey21} for a dodecagonal QC example). However, it is not feasible to realize this structure with patchy particles of well-defined radii and with bonds along the edges of the tiling because 
the distance across the short diagonal of a rhomboidal tile is 0.7654 of the tile edge length and hence would lead to particle overlaps. 
Instead, we explored how to place particles at the vertices of the Ammann-Beenker tiling at different heights in $z$ so that only tile edges correspond to bonds and there are no particle overlaps. 
Our solution is presented in Fig.\ \ref{fig:ideal}. Each bond has a component in the $z$-direction of $\pm c/4$ where $c$ is the periodic repeat in that direction. The rhombi become right- or left-handed 
helices with a pitch length of $c$, whereas the squares remain as a closed circuit of bonds, but where the pairs of diagonally-opposite particles are displaced by $c/4$ (Fig.\ \ref{fig:ideal}(c)). We choose $c$ so that the distance between particles across the short diagonal of the projected rhombi 
matches the next-neighbour distance across the diagonals of the squares.

The coordination number distribution for this ideal QC is the same as for the vertices in the Ammann-Beenker tiling with local environments having from 3 to 8 neighbours. Each of the environments is a sub-environment of the 8-coordinate environment, which involves eight-equivalent bonds in a $D_{4d}$ geometry (Fig.\ \ref{fig:ideal_env}(a)). The average coordination number is four.

The above proposed decoration of the square and two rhomboidal cells is clearly compatible with quasicrystal formation in systems made up of such cells, but it is interesting to ask how such a decoration constrains the ways that these cells could potentially combine. The decoration induces an edge-matching rule between the three types of cells (Fig.\ S2). Although this differs from the edge-matching rule of the Ammann-Beenker tiling that when combined with the appropriate vertex-matching rule ensures quasiperiodicity,\cite{Katz95} this is not necessarily a problem for our approach, because, like in previous patchy-particle quasicrystals,\cite{Noya21,Noya25} entropy maximization is expected to be the driving force for any quasiperiodicity observed.

This ideal QC is significantly different from previously observed 3-dimensional octagonal QCs,\cite{Wang87,Cao88,Damasceno17} as although they can also be analysed in terms of square-rhomb tilings, the decoration of the tiles with particles is much more complex, with most models of the experimental examples being related to the $\beta$-manganese structure.\cite{Huang91,Elenius09}
However, our ideal octagonal QC does have some structural similarities to the 3-dimensional decagonal QC observed in simulations in Ref.\ \onlinecite{Damasceno17} for a multi-minimum potential in that this quasicrystal is also based on a classical quasiperiodic tiling, namely the T\"{u}bingen tiling, with particles at the vertices of the tiles and one of the tiles (the pentagon) being associated with helical configurations.

\section{Patchy-particle design}
\label{sect:patch_design}
Our basic approach to patchy-particle design is to choose the patch geometry to match the bond directions of an environment. \cite{Wilber07,Tracey19} For quasicrystals in particular, we have found that rather than having a different particle type for each environment (e.g.\ eight in the current case), the complexity of the model can be reduced by representing environments that are subsets of each other by a single patchy particle.\cite{Noya21,Noya25} A consequence of this approach is that some of the patches would not be involved in bonds in the lower-coordinate environments, and so the particle would have an interaction energy that is lower in magnitude than its possible maximum. A concern might thus be that alternative structures that were better able to utilize all the system's patches might be more stable.

In the current case, as all the lower-coordinate environments are subsets of the eight-coordinate environment, one might initially imagine that a one-component system of eight-patch particles might be most suitable. 
However, this would lead to an average of four unused patches per particle in the ideal quasicrystal and so have a significant potential for alternative structure formation.
Instead, we initially consider a two-component system that is a mixture of five- (P5) and eight-patch (P8) particles, where it is envisaged that the eight-patch particles would be used in the 6-, 7- and 8-coordinate environments and the 5-patch particles in the 3-, 4- and 5-coordinate environments. In this case, the composition of 5- to 8-patch particles that would match the ideal quasicrystal is $(7\sqrt{2}-9)/(10-7\sqrt{2}):1\approx 8.95:1$ (Table S2). 

\begin{figure*}[t]
\includegraphics[width=18cm]{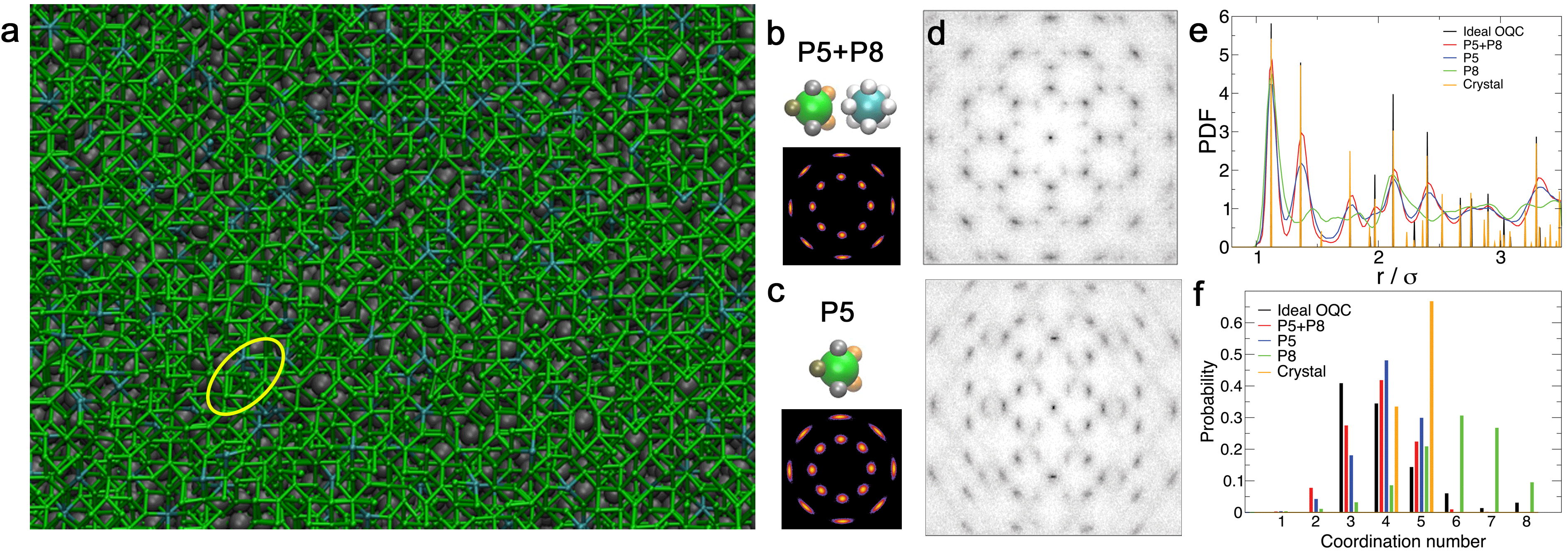}
\caption{\label{fig:OQC} 
(a) Close-up of a cut through a binary octagonal quasicrystal viewed down the 8-fold axis. Bonds are drawn between particles within $5\,\sigma_\mathrm{LJ}$ of the cut surface, and these particles are visualized as small green (P5) or cyan (P8) spheres. Particles further away from this cut plane are visualized as larger grey spheres.
The yellow ellipse highlights an ``overlapping-squares'' motif (see Fig.\ \ref{fig:crystal}(a)).
(b) BOOD for the binary quasicrystal.
(c) BOOD for a one-component quasicrystal made of 5-patch particles.
(d) Diffraction patterns of the binary octagonal QC viewed down the 8-fold axis and a 2-fold axis. 
(e) Radial distribution functions for the assembled binary, P5, and P8 clusters and the ideal OQC and $C2/c$ crystal. (f)  Coordination number distributions for the same systems as (e) (calculated for the assembled QCs using the same energy and distance criteria as for the BOODs).
}
\end{figure*}

All the patches of the 8-patch particle are equivalent by symmetry, whereas the 5-patch particle has three sets of non-equivalent patches. The specificity of the patch-patch interactions can be used to help favour the formation of the target structure. In this case, we only allow patches to interact if they would be involved in bonds in the ideal octagonal QC (the resulting interaction matrix is shown in Fig.\ \ref{fig:ideal}(e)). For example, there are 
never any bonds between 6-, 7- and 8-coordinate environments in the ideal octagonal QC, so the patches on the 8-patch particle are not self-interacting. 
Similarly, patch 2 on the 5-patch particle would never bond to itself in the ideal quasicrystal, so we also make this patch non self-interacting.

\section{Methods}
\label{sect:methods}

\subsection{Patchy-particle potential}

To simulate the patchy particles, we use a modified Lennard-Jones potential where the attractive component is modulated by angular and torsional factors such that the full attractive interaction is only obtained if interacting patches point directly at each other and the particles have the correct relative orientation.\cite{Wilber07,Wilber09b,Tracey19}  The specific form of the interactions between the patchy particles is the same as in Refs.\ \onlinecite{Wilber09b,Tracey19}. The interaction is based on a cut-and-shifted Lennard-Jones potential $V_\mathrm{LJ}^\prime$ where
\begin{equation}
V_\mathrm{LJ}^\prime(r)=
\begin{cases}
V_\mathrm{LJ}(r) - V_\mathrm{LJ}(r_\mathrm{cut}) & : r<r_\mathrm{cut}\\
0 & : r\ge r_\mathrm{cut}
\end{cases}
\end{equation}
and 
\begin{equation}
  V_\mathrm{LJ}(r)=4\varepsilon_\mathrm{LJ}\left[\left(\frac{\sigma_\mathrm{LJ}}{r}\right)^{12}-\left(\frac{\sigma_\mathrm{LJ}}{r}\right)^{6} \right]  .
\end{equation}
The patchy particle potential is described by a Lennard-Jones repulsive core and an attractive tail that is modulated by angular and torsional dependent functions:
\begin{widetext}
\begin{equation}
    V_{ij}(\mathbf{r}_{ij},\mathbf{\Omega}_i,\mathbf{\Omega}_j) = \begin{cases}
                V_{\mathrm{LJ}}^\prime(r_{ij}) & : r_{ij} < \sigma_{\mathrm{LJ}}^{\prime} \\
                V_{\mathrm{LJ}}^\prime(r_{ij}) \underset{{\mathrm{patch~pairs~}\alpha,\beta}}{\max} \left[ \varepsilon_{\alpha\beta}V_{\mathrm{ang}}(\mathbf{\hat{r}}_{ij},\mathbf{\Omega}_i,\mathbf{\Omega}_j)V_{\mathrm{tor}}(\mathbf{\hat{r}}_{ij},\mathbf{\Omega}_i,\mathbf{\Omega}_j) \right] & : r_{ij} \geq \sigma_{\mathrm{LJ}}^{\prime}
                \end{cases},
                \label{eq:V}
                \end{equation}
\end{widetext}
where $\mathbf{r}_{ij}$ is the interparticle vector,
$\alpha$ and $\beta$ are patches on particles $i$ and $j$ respectively, $\mathbf{\Omega}_i$ is the orientation of particle $i$, 
$\sigma_{\mathrm{LJ}}^{\prime}$ 
corresponds to the distance at which $V_{\mathrm{LJ}}^\prime$ passes through zero and $\varepsilon_{\alpha\beta}$ is a measure of the relative strength of the interactions between patches $\alpha$ and $\beta$.
We set the cutoff distance $r_\mathrm{cut}=2.5\,\sigma_\mathrm{LJ}$. 

The {angular} modulation term $V_\mathrm{ang}$ is a measure of how directly the patches $\alpha$ and $\beta$ point at each other, and is given by
\begin{equation}
    V_\mathrm{ang}(\mathbf{\hat{r}}_{ij},\mathbf{\Omega}_i,\mathbf{\Omega}_j) = \exp \left( -\frac{\theta_{\alpha ij}^{2}}{2\sigma_{\mathrm{ang}}^{2}} \right) \exp \left( -\frac{\theta_{\beta ji}^{2}}{2\sigma_{\mathrm{ang}}^{2}} \right) .
    \label{eq:Vang}
\end{equation}
$\theta_{\alpha ij}$ is the angle between the patch vector $\mathbf{\hat{P}}_{i}^{\alpha}$, representing the patch $\alpha$ on particle $i$, and $\mathbf{\hat{r}}_{ij}$. $\sigma_\mathrm{ang}$ is a measure of
the angular width of the patch.

The {torsional} modulation term $V_\mathrm{tor}$ describes the variation in the potential as either of the particles is rotated about the interparticle vector $\mathbf{r}_{ij}$ and is given by
\begin{equation}
    V_{\mathrm{tor}}(\mathbf{\hat{r}}_{ij},\mathbf{\Omega}_i,\mathbf{\Omega}_j)
    = \exp\left(
                - \frac{1}{2\sigma_{\mathrm{tor}}^{2}}\left[
                                \min\limits_{\phi^{\mathrm{offset}}_{\alpha\beta}}
                                \left(\phi_{\alpha\beta}-\phi^{\mathrm{offset}}_{\alpha\beta} \right)
                        \right]^{2}
        \right).
    \label{eq:Vtor}
\end{equation}
where $\phi^\mathrm{offset}$ is the preferred value of the torsional angle $\phi$.
To define the torsional angle $\phi_{\alpha\beta}$, a unique {reference vector} 
is associated with each patch.
In order to capture the symmetry of an environment, more than one equivalent offset angle can be defined,
in which case we find the minimum value of $\phi-\phi_\mathrm{offset}$ across the set of equivalent offset angles. 
The torsional interaction prevents free rotation about a patchy bond and encourages particles to bind with the correct relative orientation, thus facilitating the propagation of long-range orientational order.

As in previous work we choose $\sigma_\mathrm{tor}=2\,\sigma_\mathrm{ang}$. We also choose $\sigma_\mathrm{ang}$ 
to be sufficiently narrow to give strong directional bonding,
but not too narrow as to hinder the kinetics.
We use both $\sigma_\mathrm{ang}=0.2$ and 0.3 radians.
$\varepsilon_{\alpha\beta}$ is either 0 or 1, matching the interaction matrix in Fig.\ 1(e); as this simple choice was successful in achieving quasicrystal assembly, there was no attempt to further optimize the strengths of the interacting patches.
We use $\sigma_\mathrm{LJ}$ (the distance at which the Lennard–Jones potential is zero) as our unit of length, and the Lennard–Jones well depth $\varepsilon_\mathrm{LJ}$ as our unit of energy. 
Temperatures are given in reduced form, i.e.\ $T^*= k_B T/\varepsilon_\mathrm{LJ}$.

The details of the patchy particle designs for the binary system proposed in Section \ref{sect:patch_design} are given in Table\ S4.
The symmetry axis of the 8-patch particle is used as the reference vector for the torsional component of the potential. 
An offset angle of 180$^\circ$ ensures that the preferred relative geometry is with the eight-fold axes of the particles alligned.
The patches of the 5-patch particle are a subset of those of the 8-patch particle.

\subsection{Simulations}

The simulations were performed with a GPU-enabled Monte Carlo algorithm.\cite{Anderson13}
Our basic simulation protocol is similar to our previous work on 3-dimensional patchy-particle quasicrystals.\cite{Noya21,Tracey21,Noya25} Namely, assembly is started from a low-density fluid (with 20\,000 particles), choosing a temperature that is just sufficiently low enough to allow nucleation of the ordered phase (we generally want to avoid multiple nucleation events). 
The composition of the fluid in the binary system was chosen to match that for the ideal target quasicrystal.
Once we have grown a moderately-sized cluster (typically of order 10\,000 particles) we then place this cluster in a larger box (and with a larger reservoir of particles in the fluid phase) and continue growth. We may increase the temperature slightly in this second phase to avoid any further nucleation events. For the results reported in Section \ref{sect:Res1} and \ref{sect:Res2}, we typically ran the simulations until a cluster with close to 100\,000 particles was obtained. More details of these simulations are given in Table\ \ref{tab:sims}. 

For the simulations reported in Section \ref{sect:Res3}, which test the effects of varying potential and design parameters, instead the clusters were usually grown just until they were sufficiently large for QC formation to be 
confidently determined from the diffraction patterns and BOODs; this corresponded to at least 10\,000 particles.

\begin{table}
\caption{\label{tab:sims} 
Details of the simulations leading to the assembly of large clusters, along with the average coordination number $\langle CN\rangle$ in the interior of these clusters. For reference, $\langle CN\rangle=4$ for the ideal quasicrystal and 4\sfrac{2}{3} for the $C2/c$ crystal.
}
\begin{ruledtabular}
\begin{tabular}{ccccccc}
& & & & & max.\ \\
system 
& $\sigma_\mathrm{ang}$ 
& seed? 
&  $T_\mathrm{init}$ 
&  $T_\mathrm{growth}$ 
& cluster size 
& $\langle CN\rangle$
\\
\colrule
binary  & 0.2 & none & 0.078 & 0.078 &  80\,911 & 3.81 \\
binary  & 0.2 & ideal OQC & 0.078 & 0.079 &  79\,800 & 3.76 \\
P5     & 0.3 & none & 0.0910 & 0.0915 & 82\,330 & 4.03 \\
P5     & 0.3 & $C2/c$ crystal & 0.0910 & 0.0920 & 89\,839 & 4.03 \\
P8     & 0.3 & none & 0.11 & 0.12 & 101\,365 & 5.93 
\end{tabular}
\end{ruledtabular}
\end{table}

In the analysis of the simulation configurations we define two particles to be bonded if the distance between them is shorter than 1.5\,$\sigma_\mathrm{LJ}$ and the interaction energy is lower than $-0.2$\,$\varepsilon_\mathrm{LJ}$. An energy criterion is included because the second neighbours, although relatively close (in the ideal OQC the separation is just 22\% longer than the nearest neighbours), do not contribute significantly to the interaction energy because the P5 particle geometry means that patches rarely point at each other along these directions (Fig.\ S6). The equivalent quantities computed using just a distance criterion are presented in Section S6.

One useful way to characterize the order in the systems is using a bond orientational order diagram (BOOD).
The BOOD is a plot of the first coordination shell of
each particle on a unit sphere, which is subsequently projected onto a plane using an area-preserving Lambert projection. We choose to perform the projection onto a plane
plane perpendicular to the 8-fold symmetry axis.

\section{Results}

\subsection{Quasicrystalline binary and P5 systems}
\label{sect:Res1}

Fig.\ \ref{fig:OQC}(a) shows a cut through a large binary cluster of P5 and P8 particles grown in our simulations. From the structure it can be clearly seen that bonds are oriented along eight equivalent directions (images of the complete cluster are shown in Fig.\ S5) and no periodicity is apparent. The former feature is further confirmed by the bond-orientational order diagram (BOOD) (Fig.\ \ref{fig:OQC}(b)) which has clear eight-fold symmetry. Similarly, the diffraction pattern (Fig.\ \ref{fig:OQC}(d)) also exhibits eight-fold symmetry and quasiperiodic character, e.g.\ the ratio of the positions of the first two peaks in the $x$ direction is $\sqrt{2}$.
The cluster is an octagonal quasicrystal.

In Fig.\ \ref{fig:OQC}(f) we show the coordination number distribution for the assembled cluster. The average coordination number of our assembled octagonal QC is close to four (Table \ref{tab:sims}),
i.e.\ nearly the same as the ideal QC. However, there is a relative lack of high-coordinate particles compared to the ideal Ammann-Beenker structure. Only about 1\% of particles have a coordination number of six or more even though the fraction of 8-patch particles is 9\%. This feature suggests that the 8-patch particles may not be actually needed for quasicrystal formation. We therefore ran simulations of a one-component system of 5-patch particles (with the same patch-patch interaction matrix (Fig.\ \ref{fig:ideal}(e))). An octagonal quasicrystal again formed (Figs.\ S5 and S7)) that had very similar structural properties to the binary system, e.g.\ the coordination number distribution (Fig.\ \ref{fig:OQC}(f)) and radial distribution functions are very similar (Fig.\ \ref{fig:OQC}(e)).

\begin{figure}[t]
\includegraphics[width=8.5cm]{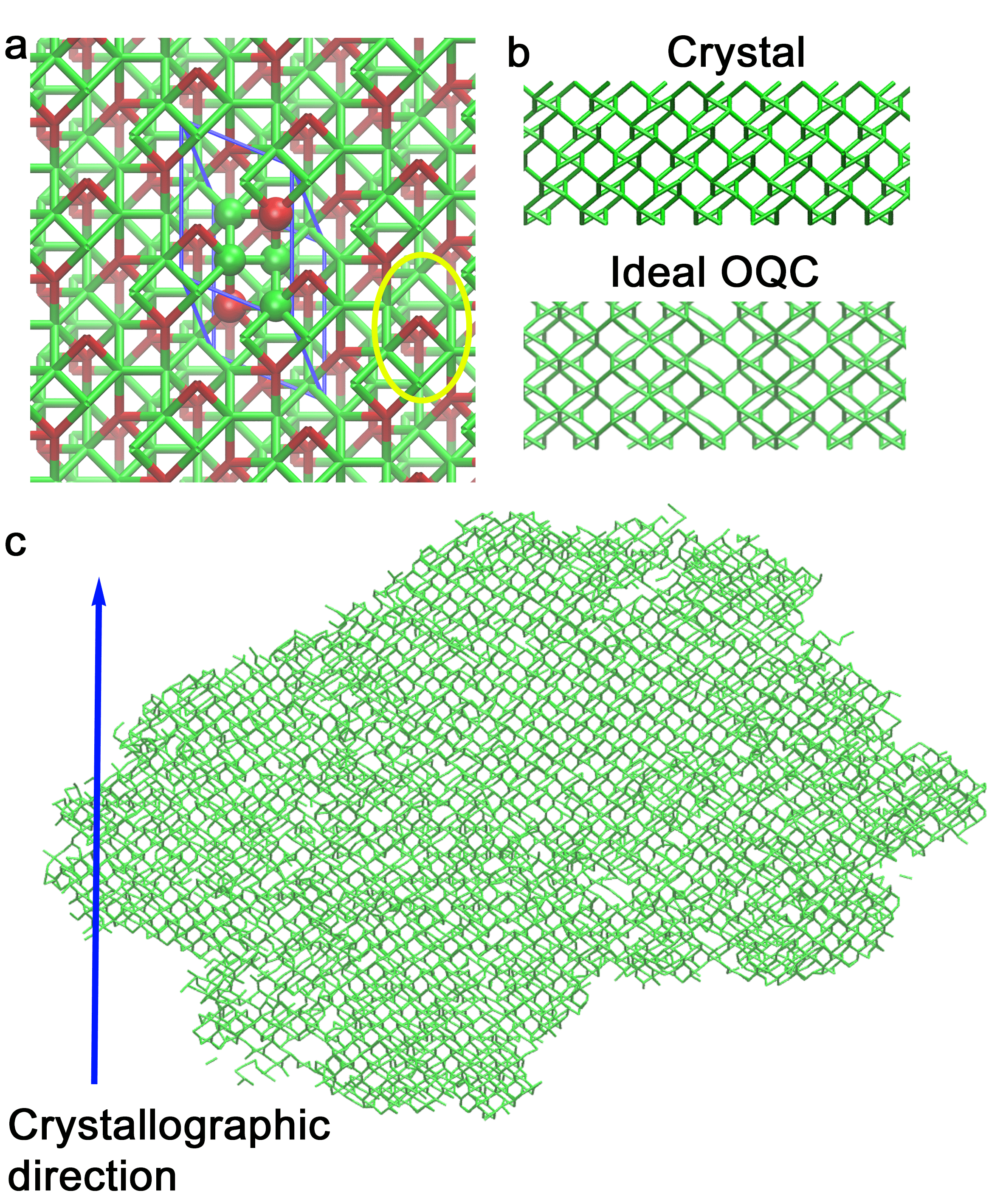}
\caption{\label{fig:crystal} (a) A slab of the 
ideal $C2/c$ crystal viewed along the pseudo 8-fold axis. The 4-coordinate environments are coloured in red and the 5-coordinate environments in green. A primitive unit cell is shown in blue (the six particles in this cell are shown as spheres). 
An ``overlapping squares'' motif is highlighted. 
The coordinates for the particles in the conventional unit cell of the crystal are given in Supplementary Table S3. The diffraction patterns for this crystal are shown in Fig.\ S9.
(b) Side views of this crystal and the ideal octagonal QC. (Further side views of the crystal are given in Fig.\ S4.)
(c) A cut of thickness 5\,$\sigma_\mathrm{LJ}$ through the P5 quasicrystal with the periodic direction vertical. 
Note that the features that look like bow ties in (b) and (c) are side views of the square motifs (Fig.\ \ref{fig:ideal}(c)).
}
\end{figure}

Crystal formation was never observed in the simulations for both these systems. In some ways it is not surprising that the periodic approximants based on the ideal QC (see Section S4) do not form as they also have a coordination number of four and so will be approximately isoenergetic with the quasicrystals but have lower entropy. We also realized that a monoclinic crystal with an average coordination number of 4\sfrac{2}{3} could be formed by the P5 particles. This crystal is illustrated in Fig.\ \ref{fig:crystal}(a). In the Ammannn-Beenker structure there are potential sites that remain unoccupied even though they are not forbidden by overlaps with the particles above and below. Exploitation of these sites leads to the higher density (Fig.\ S8(b)) and high coordination number (there are now no three-coordinate particles) of this crystal. 
Although the structure is still made up of the square and rhomboidal motifs present in the ideal octagonal QC (Fig.\ \ref{fig:ideal}(c)), the occupation of this additional site means that when viewed down the pseudo eight-fold axis, squares at different heights now overlap; this feature is highlighted in Fig.\ \ref{fig:crystal}(a). 
The similarity to the ideal octagonal quasicrystal is also seen in the radial distribution function (Fig.\ \ref{fig:OQC}(f)). The crystal also possesses the same 16 bond directions as the ideal octagonal QC, however the eight-fold symmetry is broken because one set of eight directions is 4/3 times more likely than the other. 
Furthermore, like the ideal octagonal quasicrystal, the structure consists of planes of particles perpendicular to the pseudo-eight-fold axis that are separated by $c/4$. However, as this axis is not parallel to a lattice vector of the crystal, the positions of the particles in these planes are shifted with respect to each other when viewed down this axis. 

Due to the larger coordination number, the crystal is significantly lower in energy than the octagonal quasicrystals, raising the possibility that the observed quasicrystals might just be kinetic products. In the absence of spontaneous transitions between crystals and quasicrystals,\cite{Reinhardt13b,Je21} ascertaining whether quasicrystals are thermodynamically stable or just metastable is a difficult challenge \cite{Fayen24} and one we leave to future work. 
Interestingly, when a growth simulation is started from a 279-particle crystalline seed 
the cluster that results has octagonal QC character (Fig.\ S7 and S11) and, aside from the seed, a local structure that is essentially identical to the unseeded P5 system. The crystal thus is able to template the growth of the quasicrystal. 
The preference for the quasicrystal is presumably because it has a higher growth rate, which in turn is because the greater configurational entropy inherent to the quasicrystal means there are more ways that adsorbing particles can be added that are consistent with the growth of the quasicrystalline phase.

Similarly, we also considered the growth of the binary system around a seed of the ideal octagonal quasicrystal. Unsurprisingly, this also resulted in an octagonal quasicrystal, again with negligible difference from the quasicrystal produced by homogeneous nucleation in the binary system (Fig.\ S7 and S11).

Examining the structure of the assembled quasicrystals, one can see features expected from the Ammann-Beeenker structure such as squares and thin rhombi, but one also sees additional motifs, such as the overlapping squares mentioned above (Fig.\ \ref{fig:OQC}(a) and Fig.\ \ref{fig:crystal}(a)). The side-view of the QC in Fig.\ \ref{fig:crystal}(c) (see Fig.\ S5 for a similar cut of the binary system) clearly shows that the order in the periodic direction has greater local similarity to the crystal than the ideal octagonal QC (Fig.\ \ref{fig:crystal}(b)). This helps to explain why the quasicrystals are able to match the coordination number of the ideal QC even though they possess considerable disorder. The additional motifs also provide a mechanism to introduce differences between the 2D quasicrystalline layers while maintaining inter-layer bonding. This is important as it has been suggested that 3D axial quasicrystals can only become stable due to their greater entropy if there is disorder in the periodic direction.\cite{Burkov91,Henley00} In a stacking of identical quasicrystalline layers (as in the ideal QC of Fig.\ \ref{fig:ideal}(a)) the random tiling entropy would only scale with the 2D area of the quasicrystal and so never be able to overcome an energy difference with respect to the most stable crystal that would scale with the volume.

\begin{figure}
\includegraphics[width=8.4cm]{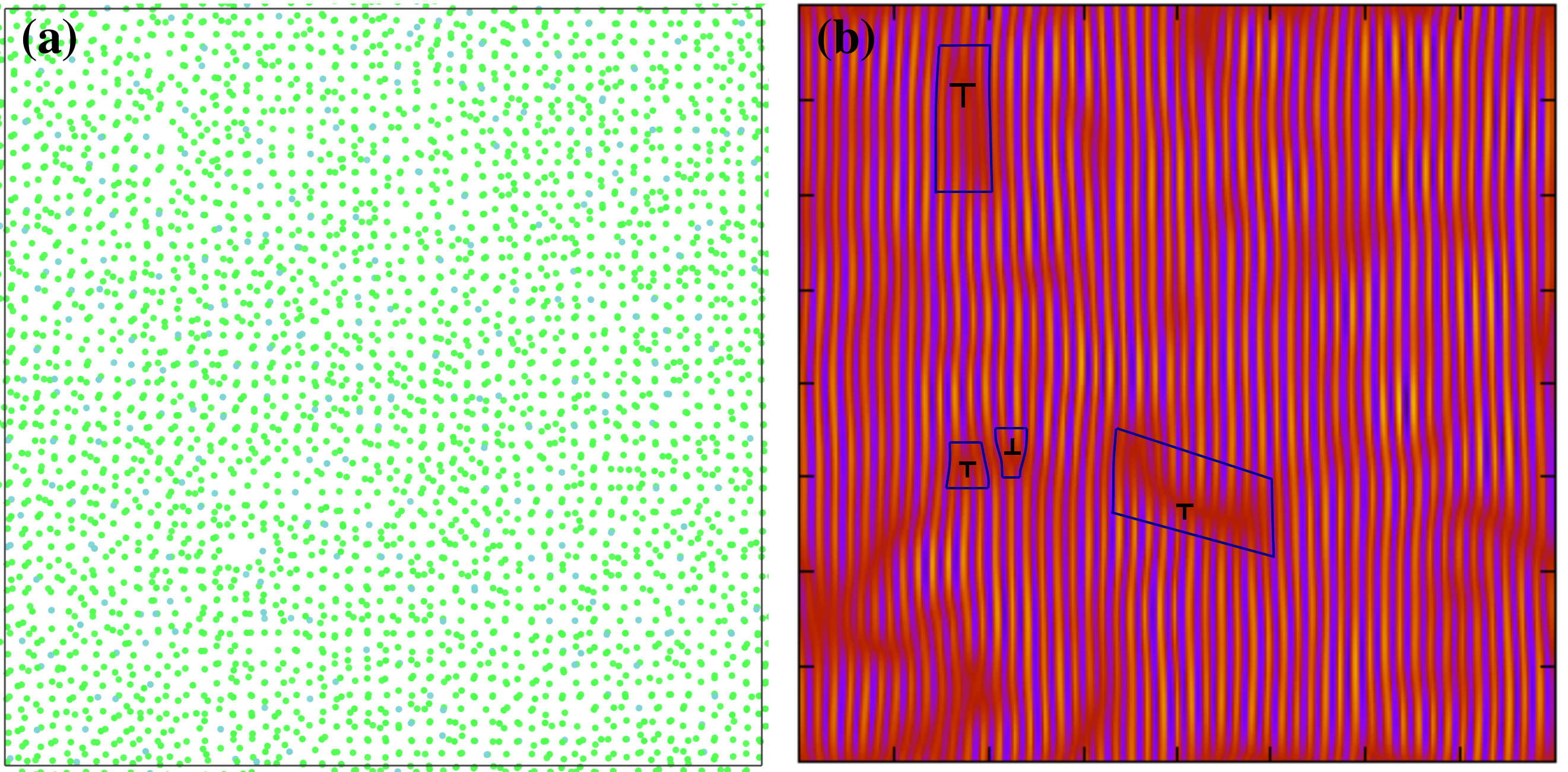}
\caption{\label{fig:dislocation} 
(a) A slab of dimensions of $20\times 20\times 4$ (in units of $\sigma_\mathrm{LJ}$) from the binary quasicrystal with the particles represented as points (P5: green; P8: cyan). If viewed at a low angle, it becomes easier to see the lines of points and hence to spot the dislocations.
(b) Inverse Fourier-transformed image of two equal and opposite diffraction spots in the first intense ring of the Fourier transform of the above slab. 
Edge dislocations are indicated by a `T'. Circuits around the dislocations are drawn in blue. If these are followed the number of lines on one side of the dislocation will be different from the other. 
(Plots for the three other pairs of equivalent diffraction spots are shown in Fig.\ S10 and enable further dislocations to be identified.) 
}
\end{figure}

A detailed examination of the configurations of the assembled quasicrystals for the binary and one-component 5-patch systems also reveal features (e.g.\ lines of particles that terminate and the bending of lines of particles near to the termination point) that are suggestive of edge dislocations in the quasicrystalline planes.
However, such features are harder to interpret with confidence for quasicrystals than for periodic crystals just by visual inspection. To obtain a more definitive identification of the presence of dislocations, we follow the approach introduced in Refs.\ \onlinecite{Korkidi13,Sandbrink14}. In this approach, two peaks at $\mathbf{q}$ and $-\mathbf{q}$ in the diffraction pattern corresponding to planes of the appropriate orientation are selected and then an inverse Fourier transform performed with the rest of the pattern masked. 
If no dislocations are present, a series of parallel lines will be obtained. However, a dislocation is revealed by the termination of one of these lines. More specifically, if a circuit is drawn around the dislocation, the number of lines on either side of the dislocation will differ by 1.

Fig.\ \ref{fig:dislocation}(b) shows such an image for one of the four equivalent directions in the binary octagonal quasicrystal generated from a thin (along the periodic direction) slab. Edge dislocations can be clearly located in the image. By contrast, we never observed any dislocations associated with the periodic direction. This approach also confirms the presence of edge dislocations in the one-component 5-patch octagonal QC.

An edge dislocation is a line defect in a 3D periodic crystal, but a point defect in a 2D periodic crystal. In the limit that the structure within each successive quasicrystalline plane is uncorrelated, the dislocations would just be point defects specific to that plane. In our quasicrystals there are correlations between successive planes in order to facilitate interplane bonding, but, as noted above, the range of this order is relatively short. This disorder means that one can only follow the path of an edge dislocation in the quasicrystal over a relatively short distance, rather than through the whole structure until it exits the surface or forms a complete loop as is typical in periodic crystals.

It is interesting to ask why these dislocations seem so common in the current octagonal systems, but are completely absent in the patchy-particle icosahedral quasicrystals that we have previously studied.\cite{Noya21,Noya25}
It may reflect that the icosahedral QC is quasiperiodic in all three dimensions,
whereas the competition between the different ways of propagating the order in the periodic direction for the octagonal QCs likely play a key role. In particular, domains of different type of order in the periodic direction are visible in the vertical cuts through the assembled quasicrystals (Fig.\ \ref{fig:crystal}(c) and Fig.\ S5) that correspond to the different potential orientations of the crystal-like ordering (Fig.\ S4). Where these meet, edge dislocations may arise.
Given their ubiquity, one would presume that in the current octagonal systems the free-energetic cost of the dislocations must be relatively low.
This may well mean that they are thermodynamically stable, rather than just being a consequence of the kinetics of growth, and so, contribute to the entropic stabilization of the QC phase.

One way to potentially assess the quality of the quasiperiodic order is to measure the phason strain (it is zero for a perfectly quasiperiodic structure).\cite{Engel15} 
This analysis requires a ``lifting'' of the particle coordinates to 5 dimensions in a reverse of the ``cut-and-project'' process that can be used to generate an ideal 3D octagonal quasicrystal. However, in the current examples, this mapping is ill-defined due to the ubiquitous presence of edge dislocations in the quasicrystalline planes. 

\begin{figure*}[t]
\includegraphics[width=16cm]{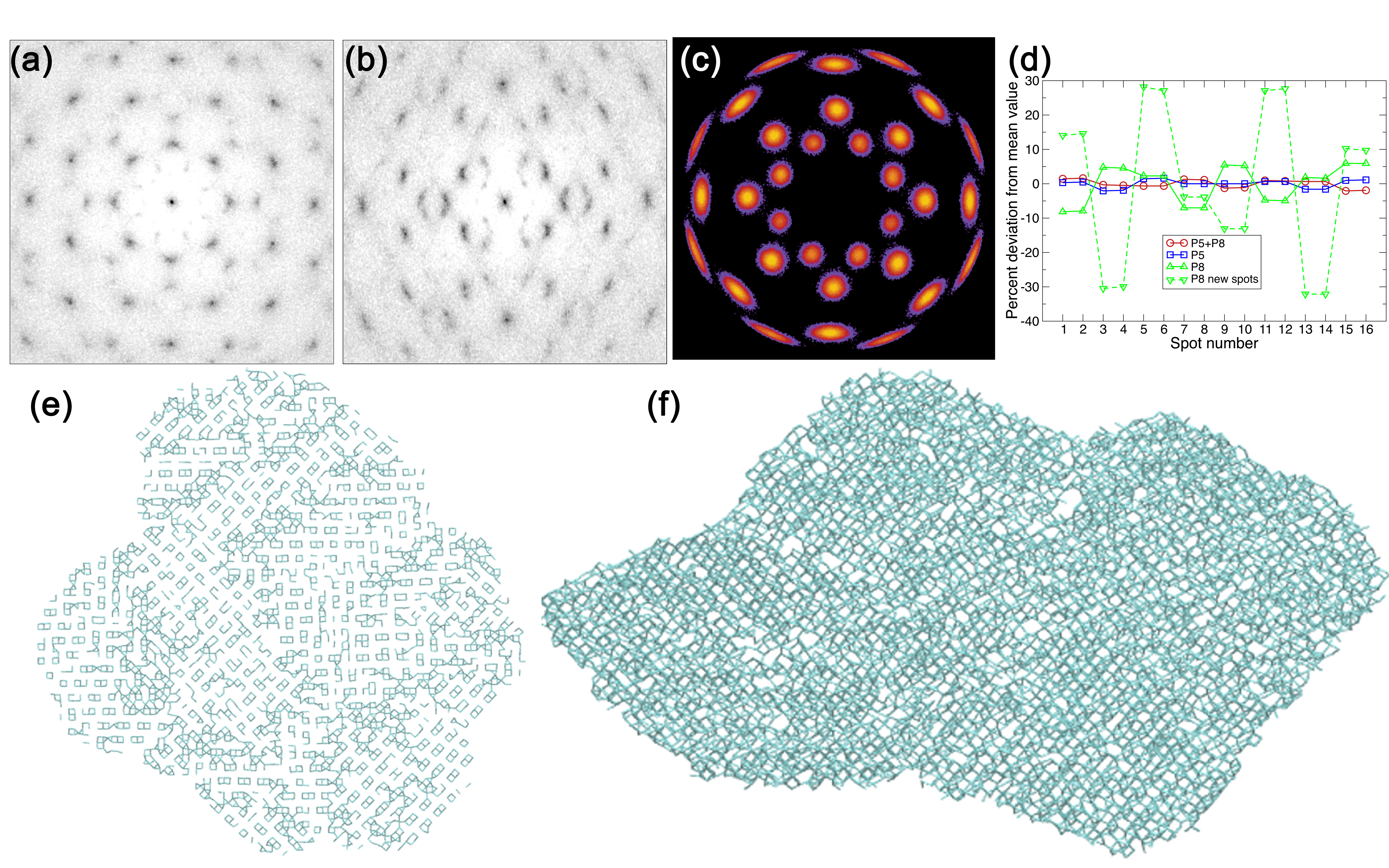}
\caption{\label{fig:1componentp8} 
(a-b) Diffraction pattern of a 101\,365 particle cluster in a one-component eight-patch system viewed (a) down the axis of approximate eight-fold symmetry and (b) perpendicular to it. (c) BOOD where bonds are defined when two particles are closer than 1.5\,$\sigma_\mathrm{LJ}$ and have an energy lower than $-0.2\,\varepsilon_\mathrm{LJ}$.  
(d) For each spot in the BOOD, the percentage deviation from the mean (averaged over symmetry-equivalent spots) value of the integral under the peak for each of the assembled QCs. 
We divide the 32 spots in the P8 BOOD into two types; namely, the more intense spots (P8) that are analogous to those that appear in the P5 and binary QCs and the less intense spots that appear due to the additional structural distortion present in the P8 system (P8 new spots).
(e-f) Slices through the cluster reveal the multi-domain character of the cluster. (e) is viewed down the approximate eight-fold axis and (f) perpendicular to it. 
}
\end{figure*}

\subsection{P8 system}
\label{sect:Res2}

We also explored what happens in a one-component system of the 8-patch particles (the patches are now allowed to self-interact). The diffraction pattern and BOOD for a large cluster are shown in Fig.\ \ref{fig:1componentp8} and exhibit features similar to that of an octagonal quasicrystal. However, the situation is more complex than for the previous systems. 
The assembled cluster shows increased facetting (the surfaces of the quasicrystals (e.g.\ Fig.\ \ref{fig:crystal}(c) and S5) are by contrast significantly rougher) and repetitive patterns that suggest local periodicity (Fig.\ \ref{fig:1componentp8} and S5). As is clear from the slices in Fig.\ \ref{fig:1componentp8}(e) and (f) the cluster consists of multiple crystalline domains. Due to the coherent boundaries between the domains, a structure with an overall approximate eight-fold symmetry results. 
The crystalline domains are based upon the crystal illustrated in Fig.\ \ref{fig:crystal}(a) but that has deformed to allow the additional patches that would not otherwise be involved in bonding to form bonds with some of the relatively close second neighbours (in the ideal octagonal QC the second neighbours are at a distance of only 1.2156 times the nearest neighbours (Fig.\ \ref{fig:OQC}(e)). These extra interactions lead to the additional rings of eight smaller peaks in the BOOD that are not present in the previous systems and that occur closer to the symmetry axis. Other consequences of this change in local structure are a larger average coordination number (nearly 6 (Table \ref{tab:sims})), a shorter repeat in the periodic direction (Fig.\ S7) and differences in the radial distribution function.
In particular, there is now no longer a peak close to $1.36\,\sigma_\mathrm{LJ}$ because the distortion causes some of the second neighbours to be incorporated into the first peak  and the others to be pushed to larger separations (Fig.\ \ref{fig:OQC}(e)).

If one looks closely at the set of smaller peaks in the BOOD associated with the new patchy bonds, one can notice that the peaks are not all of the same intensity, i.e.\ the system does not have perfect 8-fold order. Integrating the area under the peaks confirms this conclusion and provides a more quantitative measure of the deviation from perfect symmetry (Fig.\ \ref{fig:1componentp8}(d)). The integrated area associated with the weakest peaks in this set are nearly half that for the strongest. Systematic intensity deviations are also present in the other sets of peaks but these are significantly smaller in magnitude. By contrast, the intensity of the BOOD peaks for the binary and 5-patch systems are essentially identical aside from a small amount of statistical noise, further confirming their eight-fold symmetry.
One interesting question is whether close to perfect 8-fold order might be restored for larger P8 clusters if the cluster size is much larger than the domain size of the crystallites. However, the large simulations that would be required are beyond the scope of this work.

\subsection{Effects of potential and particle parameters}
\label{sect:Res3}

An important question, particular when considering experimental realization, is how sensitive is the observed behaviour to the potential and particle design parameters. Here, we begin to address this issue by considering some of the effects of patch specificity, patch width, the presence of torsions and the patch positions.

In all the simulations of the one-component 5-patch system discussed so far, the specificity of the patch-patch interactions has been that shown in Fig.\ \ref{fig:ideal}(e); i.e.\ patch 2 cannot interact with itself. We also considered a simpler system where all the patches of the P5 particles could interact with any other; such a system might be easier to realize experimentally. We found that removal of this constraint had little effect on the assembly and an octagonal QC again resulted.

We also tested how sensitive the propensity of these systems to form an octagonal quasicrystal is on patch width. The effects of the patch width on the self-assembly of a variety of classes of ordered structures, be they symmetric complexes,\cite{Wilber07,Wilber09b} crystals \cite{Doye07} or quasicrystals,\cite{Noya21} have been previously characterized. If a structure can exactly match the patch geometry, the general expectation is that the structure will remain stable as the patch width decreases but the kinetics of assembly slows down, whereas as the patch width increases there will come a point at which the potential is insufficiently directional to favour the target structure. To locate the latter limit we considered increments of 0.1 in $\sigma_\mathrm{ang}$. In the binary and the standard P5 system $\sigma_\mathrm{ang}=0.4$ radians was the last value at which structures with eight-fold order assembled, albeit with noticeably greater disorder.
Interestingly, for the P5 variant with lower specificity eight-fold order in $x$ and $y$ was lost at $\sigma_\mathrm{ang}=0.4$ but layering was still seen in $z$; the local structure was quite similar to the octagonal QC but long-range orientational order was absent. These results are consistent with the idea that systems with greater interaction specificity can tolerate greater angular flexibility in their interactions.

As noted in Section \ref{sect:Res2}, the one-component P8 systems increase their coordination number by adopting a structure in which the patches are no longer able to perfectly point at each other. The energetic penalty for this deviation from perfect alignment will increase as the patch width decreases. Indeed, at $\sigma_\mathrm{ang}=0.2$ the system instead forms planar sheets of puckered squares. These are able to achieve a coordination number of 4, whilst having significant entropy associated with the fluctuations of the sheets. The latter is the likely reason that these 2-dimensional assemblies are observed rather than one of the 3-dimensional structures with similar coordination number. Note that the patch geometry of the P5 particles is not compatible with the pattern of bonding in these sheets and so this alternative assembly is only relevant to the P8 systems.

In all the simulations described so far, the patchy-particle potential has involved a torsional component (Eq.\ \ref{eq:Vtor}) that favours interacting particles to adopt relative orientations that match the target structure, namely with the reference vectors for the interacting patches anti-parallel (Table S4). For some of the potential approaches to realizing patchy particles, it may be hard to generate a torsional component to the potential. Therefore, we also considered the assembly behaviour with the torsional component absent. However, unlike some of the systems that form icosahedral QCs,\cite{Noya21,Noya25} it was never feasible to grow any octagonal QCs without the torsional interactions present. Instead, disordered configurations always resulted. The torsional component of the potential must sufficiently disfavour otherwise feasible alternative configurations that octagonal QC formation results.

\begin{figure}[t]
\includegraphics[width=8.5cm]{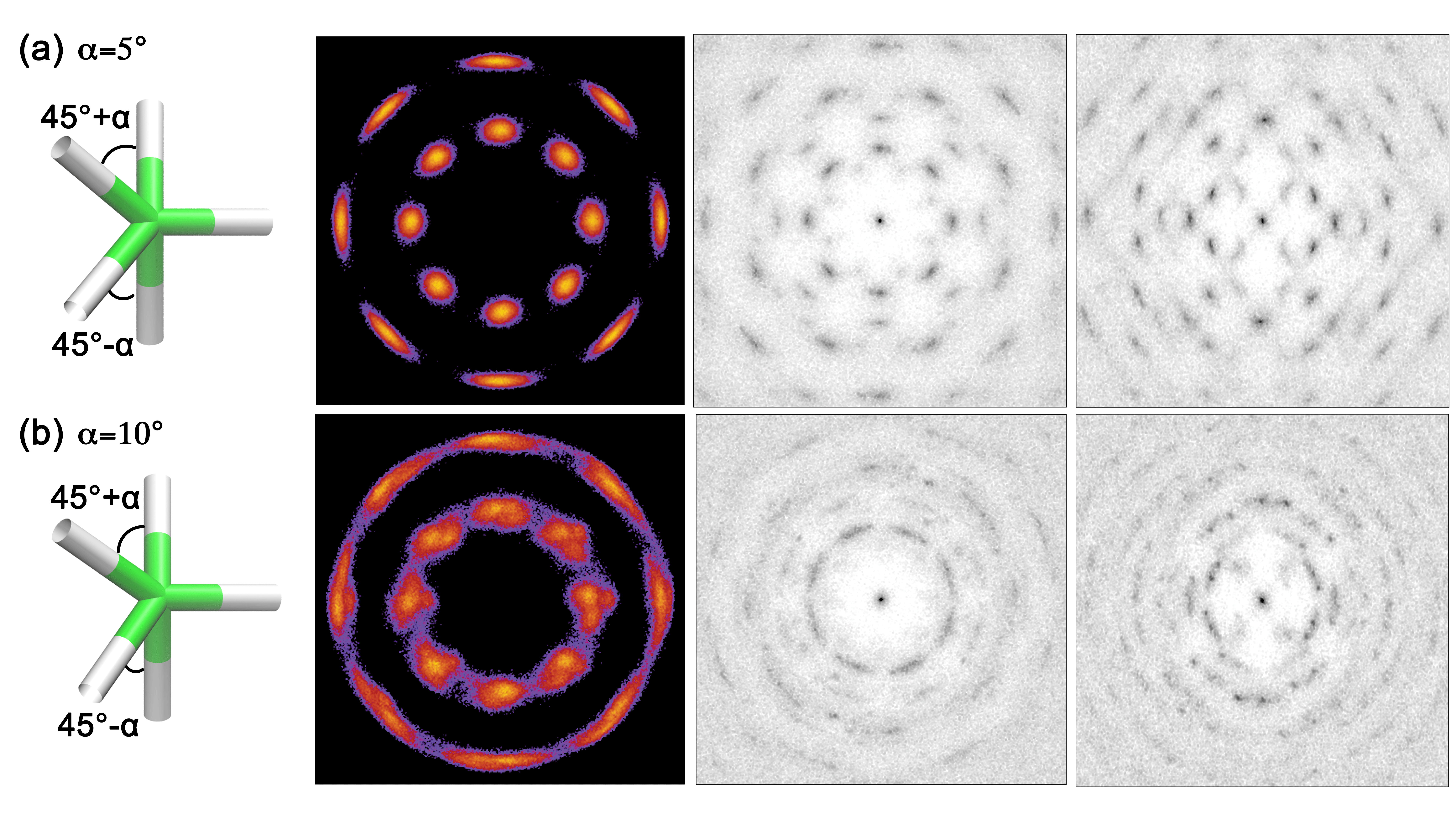}
\caption{\label{fig:chiralp5} BOOD and diffraction patterns of clusters of about 75\,000 particles assembled from one-component chiral P5 particles (in which all patches are allowed to interact among themselves) obtained by (a) a 5$^{\circ}$ and (b) a 10$^{\circ}$ rotation between the upward- and downward- facing patches.}
\end{figure}

Finally, we also considered the effects of deviations in the patch geometry. The specific deviation was an additional rotation $\alpha$ between the upward- and downward-facing patches. For the P8 particle, such a distortion would reduce its symmetry from $D_{4d}$ to $D_4$ and for the P5 particles from $C_s$ to $C_1$, thus making both particles chiral. We note that patchy particles made from proteins or DNA cannot possess any mirror symmetry. We specifically considered the effects of this distortion on systems of P5 particles. For $\alpha$=5$^\circ$ the system is still able to assemble into an octagonal QC (Fig.\ \ref{fig:chiralp5}(a)). Interestingly, the BOOD retains its $D_{4d}$ symmetry albeit with slightly wider peaks than for the undistorted P5 system. Although the particles are chiral, they adapt themselves to form a structure that is on average achiral. We note that for the ideal octagonal QC based on the Ammann-Beenker tiling 
it is clear that this structure cannot coherently distort to accommodate this change in particle geometry. 

By contrast at $\alpha$=10$^\circ$ the assembled cluster has begun to lose its global octagonal order. Although the local structure of the cluster is similar, the overall structure seems to have split into domains that are imperfectly aligned with each other (Fig.\ S12); this is reflected in the BOOD where the eight peaks are not only much wider but seem to consist of multiple sub-peaks that presumably originate from these different domains (Fig.\ \ref{fig:chiralp5}(b)). These changes are likely a reflection of the greater energetic cost associated with the mis-alignment between the patch geometry and the structure of the octagonal QC.

\section{Conclusions}

Here, we have designed patchy particles that we have shown to be capable of assembling into a 3-dimensional octagonal quasicrystal. Although the patchy-particle designs stemmed from an ideal 3-dimensional octagonal quasicrystal based on the Ammann-Beenker tiling, the structures of the assembled octagonal quasicrystals were subtly different, having a narrower coordination number distribution and a local structure, particularly in the periodic direction, that more resembled an alternative crystal structure. The former meant that the 8-patch particles that were part of the original particle design were not necessary for quasicrystal formation and, so, quasicrystals could be formed from a one-component system of 5-patch particles.
It is noteworthy that formation of the $C2/c$ crystal was never observed in any of the simulations of these two systems, even though it will be significantly lower in free energy than the quasicrystal at sufficiently low temperature. Furthermore, although local motifs similar to that in the crystal are observed, they only extend over short length scales in the quasicrystalline planes. 

The current paper is part of a programme of research seeking to understand to what extent particles with directional interactions can be used to direct the formation of 3-dimensional quasicrystals. The longer term goal is that this would then facilitate the expansion of both the types of systems for which quasicrystals can be achieved experimentally and the range of symmetries they exhibit. The current results have thus increased the symmetries of 3D patchy-particle QCs beyond just the dodecagonal \cite{Tracey21} and icosahedral \cite{Noya21,Noya25} examples that have been developed previously to now include octagonal QCs. This is particularly significant as the number of octagonal quasicrystals observed experimentally is relatively limited and it is not clear if any are actually thermodynamically stable. Furthermore, the structure of the current octagonal quasicrystals are quite different from those previously observed in experiments\cite{Wang87,Cao88}  or simulations.\cite{Damasceno17}

The driver of the quasiperiodicity is likely to be entropy maximization, like in random tiling models of quasicrystals,\cite{Henley91} but where the sources of entropy are more extensive.
Whether the octagonal QCs discovered here are thermodynamically stable (i.e.\ lowest in free energy over a certain temperature range) or just a kinetic product is also an important, but challenging, question.
Aspects of our results offer encouragement for the idea that the current octagonal QCs may be thermodynamically stable. Firstly, the quasicrystals exhibit significant disorder in the periodic direction, i.e.\ they are not identical repeats of a set of quasicrystalline planes. The above is thought to be a necessary prerequisite for the entropy of axial QCs to scale with the number of particles and hence for the QCs to be able to compete thermodynamically with an energetically more stable crystal.\cite{Burkov91,Henley00} Secondly, we never see crystal formation in the binary or P5 systems. Thirdly, we observed growth of an octagonal QC from a crystal seed. Although the latter two could be the results of purely kinetic effects, such features become more likely as the thermodynamic stability of the QC with respect to the crystal increases. In particular, if the preferential growth of an octagonal QC on the crystal persists up to the melting point, then the QC is thermodynamically more stable.

The likely best way to address this question of thermodynamic stability would be by ``direct coexistence'' simulations \cite{Vega08,Smallenburg25} to determine the melting point of the quasicrystal compared to potential competing crystals. This approach requires the use of a ``slab'' geometry where the slab crosses the periodic boundaries of the cell in two of the three dimensions. One potential issue is that the formal incompatibility of these boundary conditions with the symmetry of the quasicrystal is likely to lead to defects being present in the quasicrystal,\cite{Noya25} but for a sufficiently large box the consequent effects on the melting point may be relatively minor.

Another important question is how might particles analogous to those studied here be realized. Recently, DNA origami particles have been designed to assemble into increasingly complex crystals.\cite{Tian20, Posnjak24,Liu24,Kahn25b} This has both been through DNA origami polyhedra that assemble through single strands at their vertices \cite{Tian20,Liu24,Kahn25b} and through motifs with rigid arms that enable directional bonding.\cite{Posnjak24} Given the need for torsionally-specific interactions for our particle designs to assemble, the latter approach might be more appropriate in this case, e.g. a particle with five arms extending from a central square anti-prism. Given the impressive recent advances in protein design algorithms,\cite{Winnifrith24} proteins might provide another possibility.\cite{Wang22} For example, an approach to generate proteins that can assemble into crystals with well-defined symmetries through the hierarchical assembly of symmetric clusters has recently been developed.\cite{Li23} Using such an approach analogues of the higher-symmetry P8 particles are likely to be easier to realize, however, there have been also significant recent advances in assembling complexes with programmed symmetry breaking,\cite{Gladkov24,Lee25} as well as more arbitrary multi-component complexes using standardized protein building blocks.\cite{Huddy24} A more detailed discussion of potential realization strategies is provided in Section S7.

\section*{Supplementary Material}

The supplementary material provide further details 
on (i) the ideal QC, including its higher-dimensional description, the local environments and its crystalline approximants,
(ii) the patchy-particle designs, (iii) the alternative monoclinic crystal,
(iv) additional structural characterization of the assembled QCs
and (v) potential approaches to realize experimental analogues of the patchy particle. 

\begin{acknowledgments}
We are grateful for financial support from
Agencia Estatal de Investigaci\'on through grants No. PID2020-115722GB-C21 and PID2023-151751NB-I00 funded by MCIN/AEI/10.13039/501100011033
(E.G.N),
the Special Fund for Young Researchers of Keio University, provided by the Ishii Ishibashi Kikin, and by JSPS KAKENHI Grant Number JP202222673 (A.K.).
We acknowledge the use of the University of Oxford Advanced Research Computing (ARC) facility. 
\end{acknowledgments}

\section*{Author Declarations}

\subsection*{Conflict of Interest}
The authors have no conflicts to disclose.

\section*{Data Availability}
The code used to perform the Monte Carlo simulations is available at \url{https://github.com/evanoya/MC_GPU}. 
Data associated with the project 
is available at the Oxford University Research Archive (ORA) at \url{https://dx.doi.org/10.5287/ora-eorewogw8}.

%

\end{document}


\title{Supplementary Material for ``A patchy-particle 3-dimensional octagonal quasicrystal''}

\author{Akie Kowaguchi}
\affiliation{Physical and Theoretical Chemistry Laboratory, Department of Chemistry, University of Oxford, South Parks Road, Oxford, OX1 3QZ, United Kingdom}
\affiliation{Department of Mechanical Engineering, Keio University, Yokohama 223-8522, Japan}

\author{Savan Mehta}
\author{Jonathan P. K. Doye}
\affiliation{Physical and Theoretical Chemistry Laboratory, Department of Chemistry, University of Oxford, South Parks Road, Oxford, OX1 3QZ, United Kingdom}

\author{Eva G. Noya}
\affiliation{Instituto de Qu\'{i}mica F\'{i}sica Blas Cabrera, Consejo Superior de Investigaciones Cient\'{i}ficas, CSIC, Calle Serrano 119, 28006 Madrid, Spain}

\date{\today}

\maketitle

\setcounter{figure}{0}
 \makeatletter
 \renewcommand{\thefigure}{S\@arabic\c@figure}
 \setcounter{equation}{0}
 \renewcommand{\theequation}{S\@arabic\c@equation}
 \setcounter{table}{0}
 \renewcommand{\thetable}{S\@arabic\c@table}
 \setcounter{section}{0}
 \renewcommand{\thesection}{S\@arabic\c@section}

\section{Higher-dimensional description}

The 2D Ammann-Beenker tiling can be obtained by applying the cut-and-project method to a 4D hypercubic lattice. This space can be divided into two orthogonal sub-spaces, called the parallel (or physical) and perpendicular spaces. The vertices of the tiling are obtained by projecting those 4D lattice points that lie within the canonical 
occupation domain in perpendicular space
onto the physical space. The canonical occupation domain corresponds to the projection of the 4D unit cell onto the perpendicular space and is an octagon with edge length $a/\sqrt{2}$ and incircle of $a(1+\sqrt{2})/(2\sqrt{2})$ where $a$ is the hypercubic lattice constant.
The projection matrices that defines these two spaces are given by:
\begin{equation}
Q_\mathrm{par}=
\begin{bmatrix}
 \frac{1}{2} & 0 & -\frac{1}{2} & -\frac{1}{\sqrt{2}} \\
 \frac{1}{2} & \frac{1}{\sqrt{2}} & \frac{1}{2} & 0 \\
\end{bmatrix}
\end{equation}
and
\begin{equation}
Q_\mathrm{perp}=
\begin{bmatrix}
 -\frac{1}{2} & 0 & \frac{1}{2} & -\frac{1}{\sqrt{2}} \\
 \frac{1}{2} & -\frac{1}{\sqrt{2}}  & \frac{1}{2} & 0 \\
\end{bmatrix}.
\end{equation}
The edge length of the tiles is $a/\sqrt{2}$.

\begin{figure}[t]
\includegraphics[width=8.4cm]{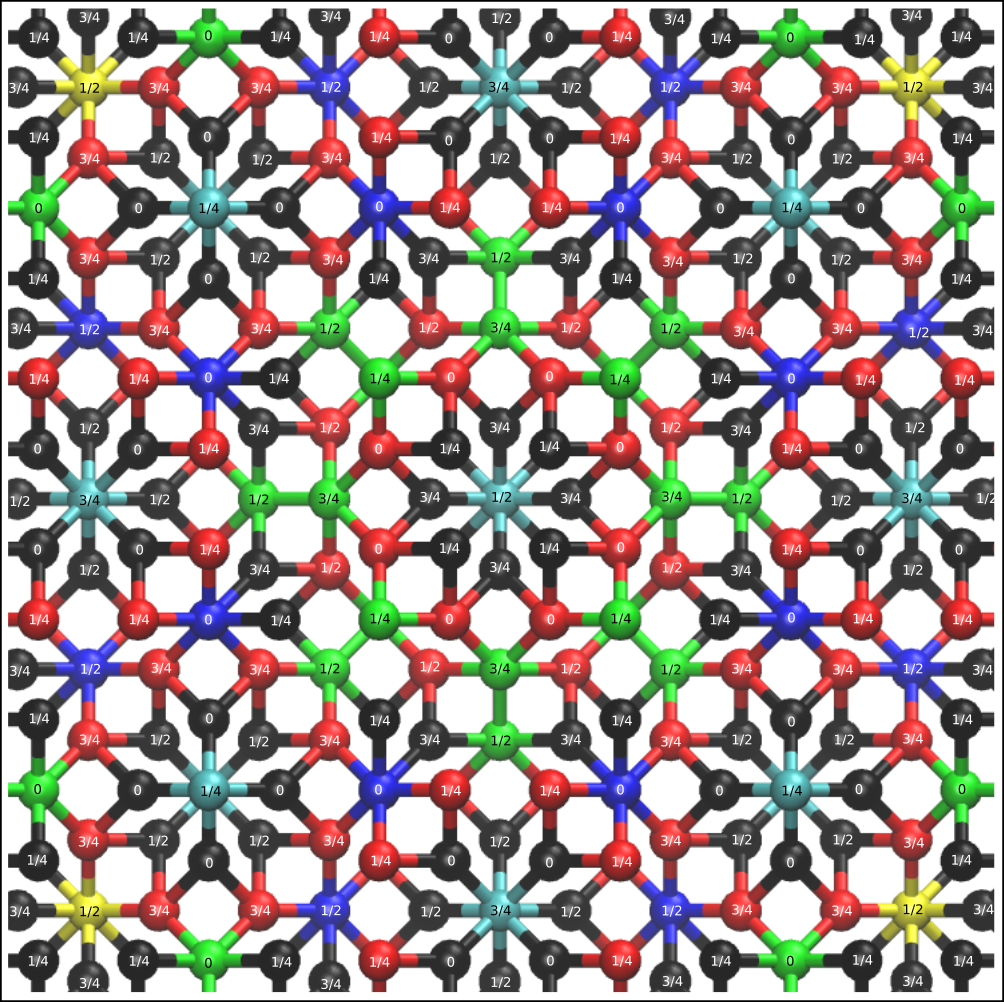}
\caption{\label{fig:ideal_zcoords} 
{Part of the ideal target octagonal quasicrystal. projected down the $c$-axis. Each particle is marked by its fractional coordinate in the $z$-direction.}
}
\end{figure}

The 3D ideal quasicrystal, which is depicted in Fig.\ 1 of the main text and Fig.\ \ref{fig:ideal_zcoords}, can be obtained in a similar way but with the addition of a fifth dimension to the higher-dimensional description. This additional coordinate simply gets directly projected onto the $z$ coordinate in the physical space. Thus, 
\begin{equation}
Q_\mathrm{par}=
\begin{bmatrix}
 \frac{1}{2} & 0 & -\frac{1}{2} & -\frac{1}{\sqrt{2}} & 0 \\
 \frac{1}{2} & \frac{1}{\sqrt{2}} & \frac{1}{2} & 0 & 0 \\
 0 & 0 & 0 & 0 & 1 \\
\end{bmatrix}
\end{equation}
and
\begin{equation}
Q_\mathrm{perp}=
\begin{bmatrix}
 -\frac{1}{2} & 0 & \frac{1}{2} & -\frac{1}{\sqrt{2}} & 0 \\
 \frac{1}{2} & -\frac{1}{\sqrt{2}}  & \frac{1}{2} & 0 & 0 \\
\end{bmatrix}
\label{eq:Qperp3D}
\end{equation}
The same octagonal occupation domain is used, but now 
for each of the points in the 4D hypercubic lattice, the additional fifth coordinate has a value of either 0, $c/4$, $c/2$ or $3c/4$. 
This leads to a doubling of the repeat length in the first four dimensions. 
(Each bond in the ideal QC corresponds to a unit step along a lattice direction in the 4D hypercubic subspace, and two consecutive bonds in the same direction lead to no change in $z$ because the change in $z$ in the first step ($\pm c/4$) is cancelled out by the second step.)
The dimensions of the 5D unit cell are $a_{5D} \times a_{5D} \times a_{5D} \times a_{5D} \times c$, where $a_{5D}=2a$. There are 16 sites per 5D hypertetragonal unit cell and these are given in Table \ref{table:5D}. These sites get projected into the physical space if they lie within the occupation domain.

\begin{table}[t]
\caption{\label{table:5D}
Sites in the 5D hypertetragonal unit cell. These are given as fractional coordinates in terms of the lattice constants $a_{5D}$ and $c$.
}
\begin{ruledtabular}
\begin{tabular}{c}
$(0,0,0,0,0)$ \\
$(1/2,0,0,0,1/4)$ \\
$(0,1/2,0,0,3/4)$ \\
$(0,0,1/2,0,1/4)$ \\
$(0,0,0,1/2,3/4$ \\
$(1/2,1/2,0,0,1/2)$ \\
$(1/2,0,1/2,0,0)$ \\
$(1/2,0,0,1/2,1/2)$ \\
$(0,1/2,1/2,0,1/2)$ \\
$(0,1/2,0,1/2,0)$ \\
$(0,0,1/2,1/2,1/2)$ \\
$(1/2,1/2,1/2,0,3/4)$ \\
$(1/2,1/2,0,1/2,1/4)$ \\
$(1/2,0,1/2,1/2,3/4)$ \\
$(0,1/2,1/2,1/2,1/4)$ \\
$(1/2,1/2,1/2,1/2,0)$
\end{tabular}
\end{ruledtabular}
\end{table}

The distance between bonded particles in the ideal QC is thus $\sqrt{a^2/2+c^2/16}$.
As we choose $c=2^{3/4}a$, this simplifies to $a\sqrt{1/2+ \sqrt{2}/8}$. If we are to choose $a$ so that this distance corresponds to the minimum in the Lennard-Jones potential (i.e.\ $2^{1/6}\sigma_\mathrm{LJ}$) then we need
\begin{equation}
a=\frac{2^{1/6}}{\sqrt{1/2+ \sqrt{2}/8}} \sigma_\mathrm{LJ} \approx 1.3644\,\sigma_\mathrm{LJ} .
\end{equation}

\section{Environment analysis}

The different coordination environments in the Ammann-Beenker tiling correspond to different zones of the octagonal domain as shown in Fig.\ 6 of Ref.\ \onlinecite{Socolar89} or Fig.\ 3.20 of Ref.\ \onlinecite{Steurer09}.
Analytic results for the fractions of particles in each environment are given in Table \ref{table:env}. These values are calculated assuming a uniform occupation of the occupation domain, i.e. they are proportional to the areas of the occupation domain corresponding to each coordination environment \cite{Baake90}.
\begin{table}[t]
\caption{\label{table:env}
Environments in the ideal quasicrystal. 
}
\begin{ruledtabular}
\begin{tabular}{ccc}
\textrm{Index/Label} &
$n_\mathrm{neigh}$ &
\textrm{fraction} \\
\colrule
1 & 3 & $\sqrt{2}-1\approx 0.4142$ \\
2 & 4 & $6-4\sqrt{2}\approx 0.3431$\\
3 & 5 & $10\sqrt{2}-14\approx 0.1421$ \\
4 & 6 & $34-24\sqrt{2}\approx 0.0589$\\
5 & 7 & $29\sqrt{2}-41\approx 0.0122$\\
6 & 8 & $17-12\sqrt{2}\approx 0.0294$\\
\end{tabular}
\end{ruledtabular}
\end{table}

\section{Matching rules}

\begin{figure}[t]
\includegraphics[width=8.4cm]{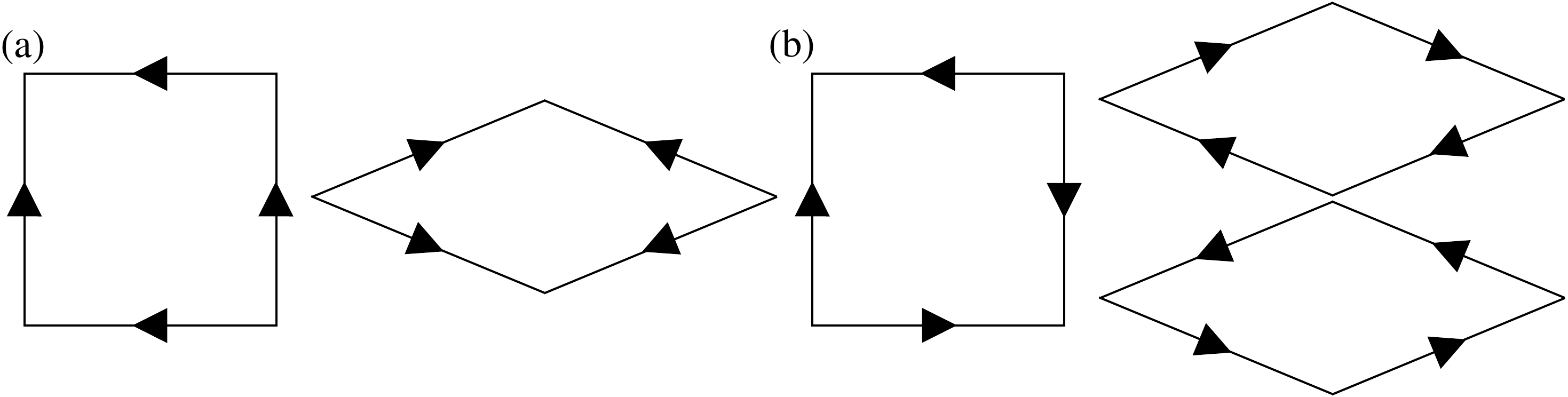}
\caption{\label{fig:matching} 
{(a) Edge-matching rules for the Ammann-Beenker tiling. (b) Effective matching
rules for the 3-dimensional ideal octagonal QC. Shared edges where the arrows point in the
same direction are allowed. In the ideal octagonal QC the effective matching rules arises
from the upward or downward direction of the bonds along the edges}
}
\end{figure}

{
A set of matching rules have been worked out for the Ammann-Beenker tiling that ensures a quasiperiodic tiling results if those matching rules are obeyed. These rules consist of constraints on the shared edge of two adjacent tiles and on the configuration of tiles around a common vertex.
These rules are usually represented by a decoration of the tiles.
Fig.\ \ref{fig:matching}(a) shows the edge rule for the Ammann-Beenker tiling in terms of a decoration of the edges of the tiles with arrows where shared edges must have the arrows pointing in the same direction. (Note this edge rule on its own (i.e.\ without the vertex rule) is insufficient to ensure quasiperiodicity \cite{Katz95}.)

An interesting question to ask is whether the particle decoration of the square and rhomboidal cells in our proposed ideal octagonal QC provides any constraints on how these cells could combine. There are and the effective edge matching rules are shown in Fig.\ \ref{fig:matching}(b). As a reminder, the decoration of the cells consists of placing particles at the vertices of the 2D projected tiling, but at heights in the periodic direction that ensure a network of bonds along the edges but with no particle overlaps. In the square cell, the four particles form a puckered square and in the rhomboidal cell the four particles form a right- or left-handed helix-like arrangement (Fig.\ 1(c) in the main text).
Each bond has a component of $\pm c/4$ in the $z$-direction. The arrows along each edge in Fig.\ \ref{fig:matching}(b) mark the upward direction.

Note that this different edge-matching rule is not a problem for quasicrystal formation in our systems, because, firstly, it does not preclude quasicrystal formation, and, secondly, the reason for quasicrystal formation is most likely entropy maximization, similar to in random tiling models of quasicrystals \cite{Henley91}.}

\section{Approximants}
Rational approximants can be derived by projection, but where a rational approximation to an irrational number, in the octagonal case either $\sqrt{2}$ or $1+\sqrt{2}$, is used in the derivation of $Q_\mathrm{perp}$. The octonacci (or Pell) sequence is
\begin{equation}
0, 1, 2, 5, 12, 29, 70, \dots
\end{equation}
where the Pell numbers are defined by the recurrence relation:
\begin{equation}
P_n=2P_{n-1}+P_{n-2} .
\end{equation}
The ratio of two successive Pell numbers $q/p=P_{n+1}/P_n$ tends to $1+\sqrt{2}$ (sometimes called the silver mean) as $n$ tends to infinity.
Therefore, $(q-p)/p$ provides a series of approximations to $\sqrt{2}$.

\begin{figure*}[t]
\includegraphics[width=18.0cm]{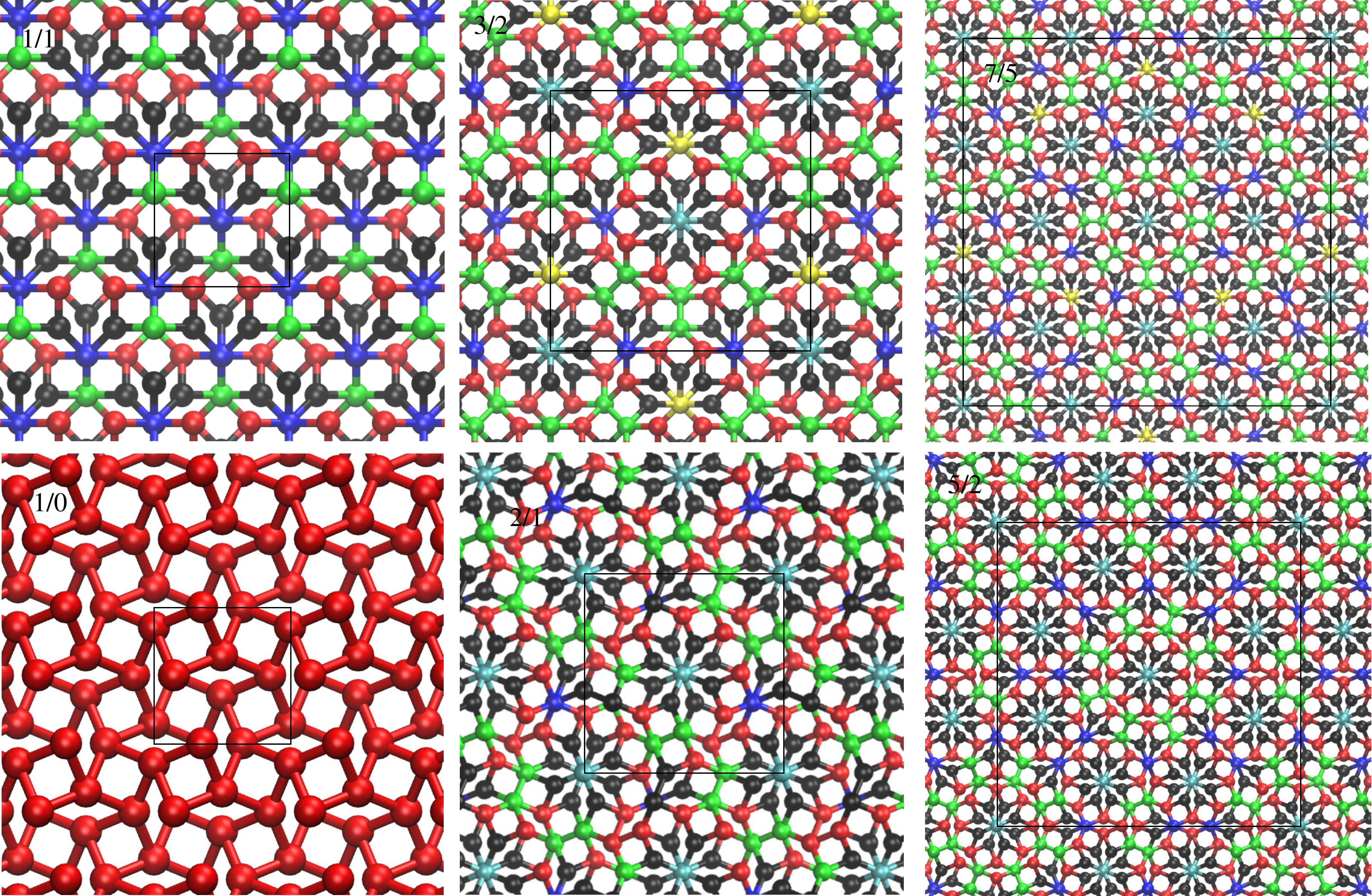}
\caption{\label{fig:approx} 
Rational approximants viewed along the $z$-axis. 
First row: series 1: 1/1, 3/2, 7/5. 
Second row: series 2: 1/0, 2/1, 5/2. 
Particles are coloured by their coordination number in the same way as in Fig.\ 1 of the main text.
The black squares correspond to the primitive 3D unit cells. The primitive unit cells for the 2D tilings are rotated by $45^\circ$ and with cell lengths a factor of $\sqrt{2}$ smaller.
}
\end{figure*}

The approach we use is to apply a shear matrix $A$ to perpendicular space, so as to reduce the perpendicular space component of appropriate 5D inter-site vectors to zero \cite{Steurer09}. (Note that in the following the 2D rational approximants to the Ammann-Beenker tilings can be obtained by ignoring the fifth dimension.) We choose 
\begin{equation}
\mathbf{r}_x=a
\begin{pmatrix}
q-p \\
0 \\
-(q-p)\\
-2p \\
0
\end{pmatrix}
\quad
\mathrm{and}
\quad
\mathbf{r}_y=a
\begin{pmatrix}
q-p \\
2p \\
q-p \\
0 \\
0
\end{pmatrix}
\end{equation}
as their parallel components are along $x$ and $y$, respectively, and their perpendicular components would be zero if $(q-p)/p=\sqrt{2}$.
Specifically,
\begin{equation}
Q\mathbf{r}_x=a
\begin{pmatrix}
(q-p)+\sqrt{2}p \\
0 \\
0 \\
-(q-p)+\sqrt{2}p \\
0 
\end{pmatrix}
\end{equation}
and
\begin{equation}
Q\mathbf{r}_y=a
\begin{pmatrix}
0 \\
(q-p)+\sqrt{2}p \\
0 \\
0 \\
(q-p)-\sqrt{2}p \\
\end{pmatrix}
\end{equation}
where
\begin{equation}
Q=
\begin{pmatrix}
Q_\mathrm{par} \\
Q_\mathrm{perp} 
\end{pmatrix}.
\end{equation}
Setting the perpendicular components of $AQ\mathbf{r}_x$ and $AQ\mathbf{r}_y$ to zero and solving
leads to
\begin{equation}
A=
\begin{pmatrix}
1 & 0 & 0 & 0 & 0 \\
0 & 1 & 0 & 0 & 0 \\
0 & 0 & 1 & 0 & 0 \\
A_{41} & 0 & 0 & 1 & 0 \\
0 & -A_{41} & 0 & 0 & 1 \\
\end{pmatrix}
\end{equation}
where
\begin{equation}
A_{41}=\frac{(q-p)-\sqrt{2}p}{(q-p)+\sqrt{2}p}.
\end{equation}
As
\begin{equation}
AQ=
\begin{pmatrix}
Q_\mathrm{par} \\
Q_\mathrm{perp}^\mathrm{approx} 
\end{pmatrix}
\end{equation}
it follows that
\begin{equation}
Q_\mathrm{perp}^\mathrm{approx}=\frac{\sqrt{2}}{(q-p)+\sqrt{2}p}
\begin{bmatrix}
 -p & 0 & p & p-q & 0 \\
 p & p-q & p & 0 & 0 \\
\end{bmatrix}.
\end{equation}
This has a similar form to Eq.\ \ref{eq:Qperp3D} but where $\sqrt{2}$ has effectively been replaced by $(q-p)/p$.
We hence label these approximants by the rational fractions 
$(q-p)/p$, i.e.\ $1/1, 3/2, 7/5, 17/12, \dots$. 

The unit cell is orthorhombic (space group: $P{mn}2_1$; note that the two-fold screw axes are perpendicular to our $z$ axis) 
with a repeat length along $x$ and $y$ of $((q-p)+\sqrt{2}p)a$.
The first three examples in this series of approximants are illustrated in the first row of Fig.\ \ref{fig:approx}.
(Note that in the case of 2D rational approximants this unit cell is centred; thus, there is a primitive cell that is oriented at $45^\circ$ degrees to the $x$ and $y$ axes and is a factor of $\sqrt{2}$ smaller. Examples have been described in Refs.\ \onlinecite{Socolar89,Duneau89,Subramaniam03}.)

In the above $\mathbf{r}_x$ and $\mathbf{r}_y$ were chosen to be along directions corresponding to two of the eight equivalent directions along which tile edges are oriented in the octagonal QCs. A second set of approximants can be derived by instead choosing to set the perpendicular components to zero in directions that bisect these directions.
We choose vectors of the form
\begin{equation}
\mathbf{r}_1=a
\begin{pmatrix}
q \\
p \\
-p\\
-q \\
0
\end{pmatrix}
\quad
\mathrm{and}
\quad
\mathbf{r}_2=a
\begin{pmatrix}
p \\
q \\
q \\
p \\
0
\end{pmatrix}
\end{equation}
that lie at $22.5^\circ$ to the $x$ and $y$ axes, respectively,
and whose perpendicular components would be zero if $q/p=1+\sqrt{2}$.

Setting the perpendicular components of $AQ\mathbf{r}_1$ and $AQ\mathbf{r}_2$ to zero and solving 
leads to
\begin{equation}
A=
\begin{pmatrix}
1 & 0 & 0 & 0 & 0 \\
0 & 1 & 0 & 0 & 0 \\
0 & 0 & 1 & 0 & 0 \\
0 & A_{42} & 0 & 1 & 0 \\
A_{42} & 0 & 0 & 0 & 1 \\
\end{pmatrix}
\end{equation}
where
\begin{equation}
A_{42}=\frac{(\sqrt{2}+1)p-q}{p+(\sqrt{2}+1)q}.
\end{equation}
Hence,
\begin{equation}
Q_\mathrm{perp}^\mathrm{approx}=\frac{1}{2}
\begin{bmatrix}
 A_{42}-1 & \sqrt{2}A_{42} & A_{42}+1 & -\sqrt{2} & 0 \\
 A_{42}+1 & -\sqrt{2} & 1-A_{42} & -\sqrt{2} A_{42} & 0 \\
\end{bmatrix}.
\end{equation}
The length of the vectors $\mathbf{r}_1$ and $\mathbf{r}_2$ in parallel space is
\begin{equation}
   q\,r_\mathrm{long} + p\, r_\mathrm{short} .
\end{equation}
where $r_\mathrm{long}=a\sqrt{1+1/\sqrt{2}}$ and $r_\mathrm{short}=a\sqrt{1-1/\sqrt{2}}$ are the distances associated with the long and short diagonals of the rhomb. 
For the 2D rational approximants these are the dimensions of the primitive unit cell. However, 
although $\mathbf{r}_1$ and $\mathbf{r}_2$ are lattice vectors in the hypercubic subspace, they are not actually inter-site vectors in the full 5 dimensions.
Instead, the primitive cell vectors for the 3D approximants are the projections in physical space of $\mathbf{r}_1+\mathbf{r}_2$ and $\mathbf{r}_1-\mathbf{r}_2$ with a cell length a factor of
$\sqrt{2}$ larger.

To differentiate these approximants from the previous set we label them by $q/p$, i.e.\ $1/0, 2/1, 5/2, 12/15, \dots$. The first three examples in this series are illustrated in the second row of Fig.\ \ref{fig:approx}. The $1/0$ approximant gives a $\beta$-Mn-like tiling of squares and rhombs (note the decoration of these tiles with particles is very different from $\beta$-Mn).
The 2D version of the $2/1$ tiling has been reported in Ref.\ \onlinecite{Socolar89}. 

\section{Alternative crystal}

The details of the alternative crystal that it is possible for the 
patchy particles to form are given in Table \ref{table:crystal}.
These coordinates allow the patches to point directly at each other along each bond.
The lattice type is monoclinic and the space group is $C2/c$.
The views depicted in Fig.\ 4 of the main text are not along the lattice directions, but are with the reference vector of the patchy particles out of the plane (Fig.\ 4(a)) and in the vertical direction (Fig.\ 4(b)) in order to be analogous to the views of the ideal and assembled quasicrystals. As well as having a higher average coordination number than the ideal octagonal QC it also has a higher density.

\begin{table}[t]
\caption{\label{table:crystal}
Wyckoff sites in the ideal $C2/c$ crystal. The distance unit is 
$\sigma_\mathrm{LJ}$. The bond lengths are set to $2^{1/6}$ (the minimum in the Lennard-Jones potential). $a=3.409$, $b=2.521$, $c=2.120$, $\alpha=\gamma=90$ and $\beta=104.923^\circ$.
}
\begin{ruledtabular}
\begin{tabular}{cccc}
Wyckoff & 
multiplicity &
coordinates &
coordination \\
site & & & number \\
\colrule
f & 8 & $(0.293,0.073,0.164)$ & 5 \\
e & 4 & $(0.000,0.220,0.250)$ & 4 \\
\end{tabular}
\end{ruledtabular}
\end{table}

\begin{figure*}
\includegraphics[width=14cm]{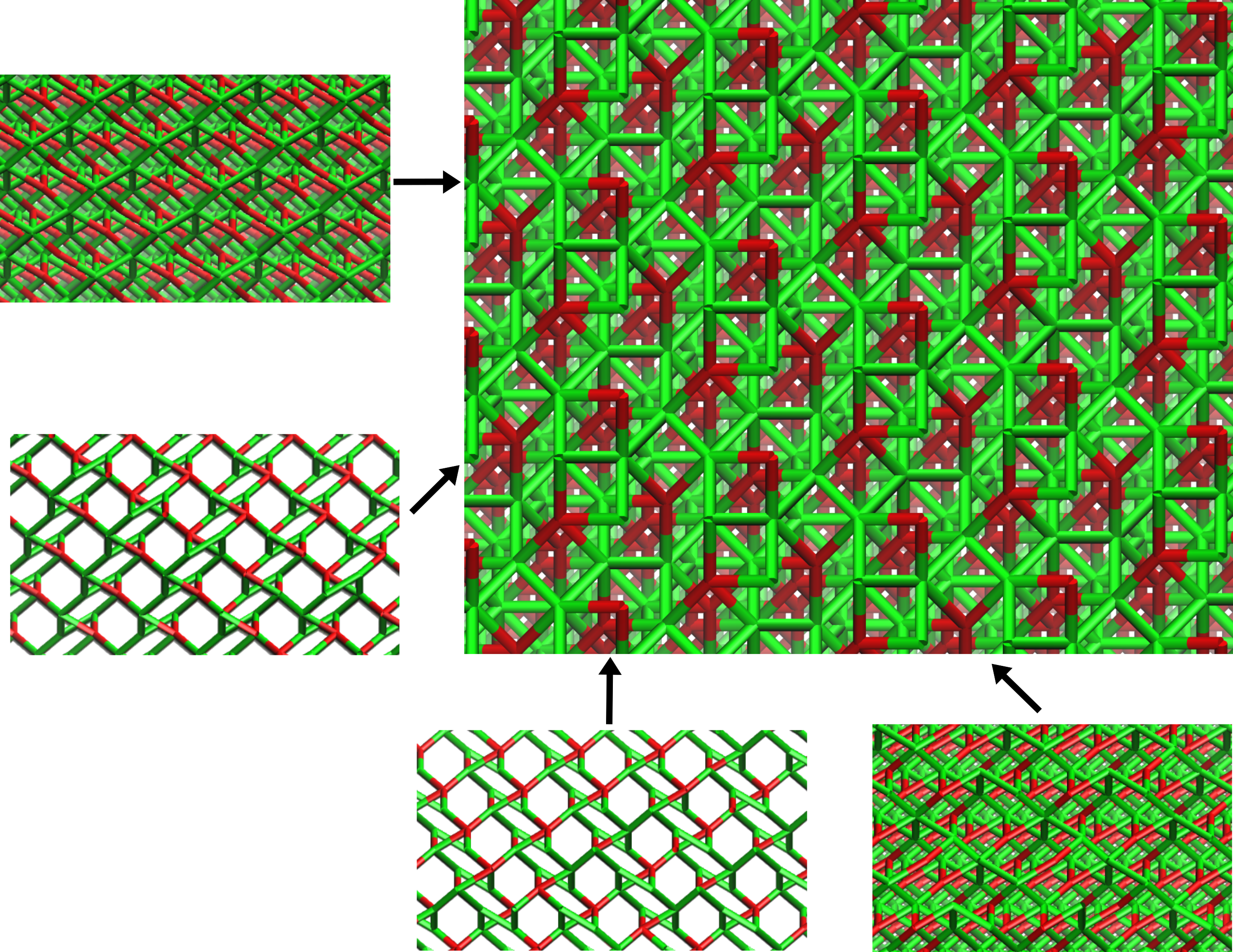}
\caption{\label{fig:crystal_side} 
{Top and four side views of the $C2/c$ crystal when the crystal is oriented so that the bonds are oriented in the same directions as in the octagonal quasicrystals. Thus, the top view is analogous to looking down the 8-fold axis of the QC and the side views to looking along one of the 2-fold axes of the QC; because the crystal does not possess eight-fold symmetry these latter views are different. The arrows indicate the viewing directions of the side views. Only in two views is the viewing direction parallel to a lattice vector of the crystal. In the others a dense network of projected bonds is observed, because, even though the same patterns of bonds is repeated at different heights in the viewing direction, they are shifted with respect to each other in the plane. As in Fig.\ 4 the 4-coordinate environments are coloured in red and the 5-coordinate environments in green.
The vertical cuts through the quasicrystals in Figs.\ 4(c) and \ref{fig:snap} (for both the viewing direction is along a 2-fold axis) show features resembling all four side views. }
}
\end{figure*}

\begin{table*}
\caption{\label{table:binary_patches}
Patchy-particle design for the binary system. The directions of the patches on the particle surface are specified by the patch unit vectors. For each patch, the reference vector and offset angles used for evaluating the torsional interactions are provided. The patches are divided into four types and the patch specificity defines the patch types with which a patch interacts. In the first column the symmetry of the particle and the colour with which it is represented in the figures is also given.
}
\begin{ruledtabular}
\begin{tabular}{ccccccc}
Particle  & Patch & Patch & Patch & Reference  & Offset & Patch\\
 type & number & type &  vector &  vector &  angles  & specificity \\
\colrule
 P5    & P$^1_A$    &  1 & $(-0.85953250, 0, 0.51108108)$ & $(0, 0, 1)$   & 180$^\circ$ &  $1,2,3,4$ \\
green & P$^2_A$    &  2 & $(0, 0.85953250, 0.51108108)$  & $(0, 0, 1)$   & 180$^\circ$ &  $1,3$ \\
$C_s$ & P$^3_A$    &  2 & $(0, -0.85953250, 0.51108108)$ &  $(0, 0, 1)$   & 180$^\circ$ &  $1,3$ \\
      & P$^4_A$    &  3 & $(0.60778126, 0.60778126, -0.51108108)$  &  $(0, 0, -1)$ & 180$^\circ$ &  $1, 2, 3, 4$ \\
      & P$^5_A$    &  3 & $(0.60778126, -0.60778126, -0.51108108)$ &  $(0, 0, -1)$ & 180$^\circ$ &  $1,2,3,4$ \\
\colrule
 P8   & P$^1_B$    &  4 & $(-0.85953250, 0, 0.51108108)$ & $(0, 0, 1)$   & 180$^\circ$ &  $1, 3$ \\
cyan & P$^2_B$    &  4 & $(0, 0.85953250, 0.51108108)$ & $(0, 0, 1)$ & 180$^\circ$  & $1, 3$ \\
$D_{4d}$ & P$^3_B$    &  4 & $(0.85953250, 0, 0.51108108)$ & $(0, 0, 1)$  & 180$^\circ$ & $1, 3$  \\
     & P$^4_B$    &  4 & $(0, -0.85953250, 0.51108108)$ & $(0, 0, 1)$ & 180$^\circ$  & $1, 3$ \\
     & P$^5_B$    &  4 & $(-0.60778126, 0.60778126, -0.51108108)$ & $(0, 0, -1)$  & 180$^\circ$ & $1, 3$ \\
     & P$^6_B$    &  4 & $(0.60778126, 0.60778126, -0.51108108)$ & $(0, 0, -1)$  & 180$^\circ$ & $1, 3$  \\
     & P$^7_B$    &  4 & $(0.60778126, -0.60778126, -0.51108108)$ &  $(0, 0, -1)$ & 180$^\circ$ & $1, 3$ \\
     & P$^8_B$    &  4 & $(-0.60778126, -0.60778126, -0.51108108)$ &  $(0, 0, -1)$ & 180$^\circ$ & $1, 3$ \\
\end{tabular}
\end{ruledtabular}
\end{table*}

\begin{table}
\caption{\label{tab:sims} 
Details of the simulations leading to the assembly of large clusters.
}
\begin{ruledtabular}
\begin{tabular}{cccccc}
system 
& $\sigma_\mathrm{ang}$ 
& seed? 
&  $T_\mathrm{init}$ 
&  $T_\mathrm{growth}$ 
& max.\ cluster size 
\\
\colrule
binary  & 0.2 & none & 0.078 & 0.078 &  80\,911 \\
binary  & 0.2 & ideal OQC & 0.078 & 0.079 &  79\,800 \\
P5     & 0.3 & none & 0.0910 & 0.0915 & 82\,330 \\
P5     & 0.3 & $C2/c$ crystal & 0.0910 & 0.0920 & 89\,839 \\
P8     & 0.3 & none & 0.11 & 0.12 & 101\,365
\end{tabular}
\end{ruledtabular}
\end{table}

\section{Further structural characterization}

\begin{figure*}
\includegraphics[width=18cm]{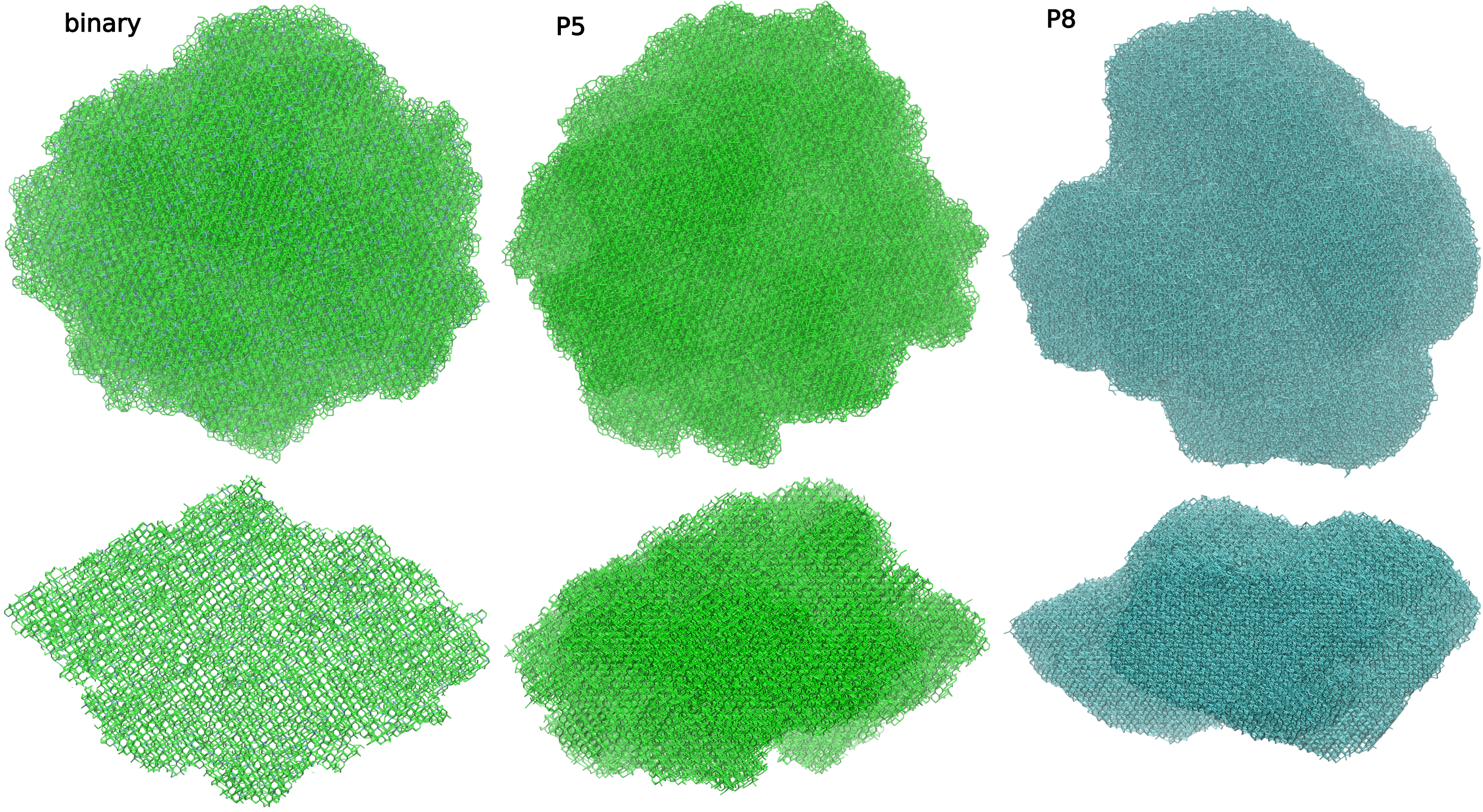}
\caption{\label{fig:snap} 
Additional snapshots of the binary, and one-component 5-patch and 8-patch clusters. The top row shows a view down the 8-fold axis and the bottom row a side view with the periodic axis vertical. For the binary system the side view is of a slice through the cluster, whereas for the other two systems, it is of the whole cluster. The greater facetting in the P8 system is evident.
}
\end{figure*}

Additional snapshots of the grown clusters are given in Fig.\ \ref{fig:snap} to supplement those in the main text. In particular, the vertical slice through the binary system again shows that the local structure {in the periodic direction} has similarities to the $C2/c$ crystal. The greater facetting of the P8 cluster is clear, reflecting its multi-crystalline character. {Somewhat more subtle is the recognition of periodic motifs. These periodic patterns are apparent in the vertical slices of the clusters and the top view of the P8 cluster, but absent from the top views (i.e.\ down the eight-fold axis) of the binary and P5 clusters, reflecting their quasiperiodicity in the planes perpendicular to the eight-fold axis.}

\begin{figure*}
\includegraphics[width=18cm]{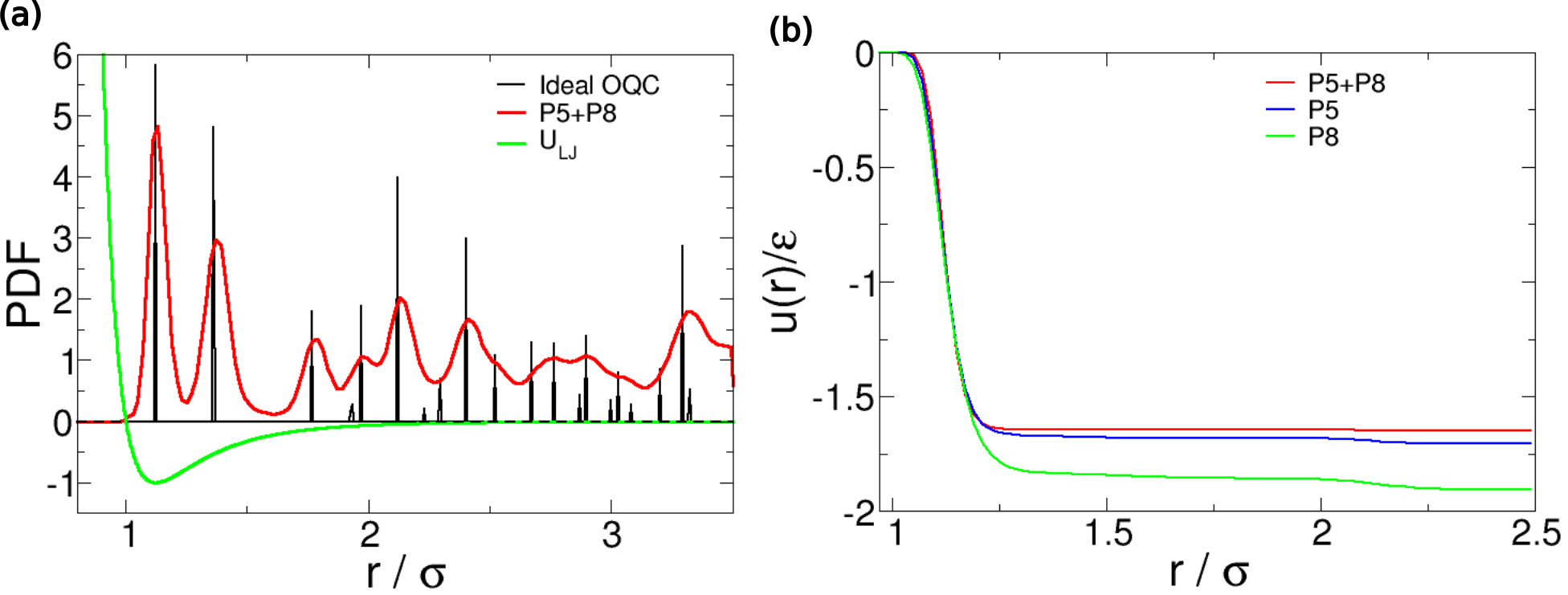}
\caption{\label{fig:interactions} 
(a) A joint plot showing both the radial distribution functions for the ideal octagonal QC and assembled binary quasicrystal, and the Lennard-Jones potential.
(b) The contribution to the potential energy per particle from pairs with distance less than $r$.
}
\end{figure*}

{Our general expectation when studying patchy-particle systems is that the nearest-neighbour pairs will dominate the interaction energy, not only because of the relatively short-ranged character of the $1/r^6$ attraction of the Lennard-Jones potential, but because next neighbours are likely to have a small value of the $V_\mathrm{ang}V_\mathrm{tor}$ term that modulates the potential (Eq.\ 3). The structure of the ideal octagonal QC is such that the second neighbours have an unusually short separation that for an isotropic Lennard-Jones potential would give such a pair an energy that is 52\% of that at the minimum (Fig.\ \ref{fig:interactions}(a)). However, Fig.\ \ref{fig:interactions}(b) shows that the vast majority of the interaction energy in our patchy-particle systems comes from the first neighbours.}

\begin{figure*}
\includegraphics[width=18cm]{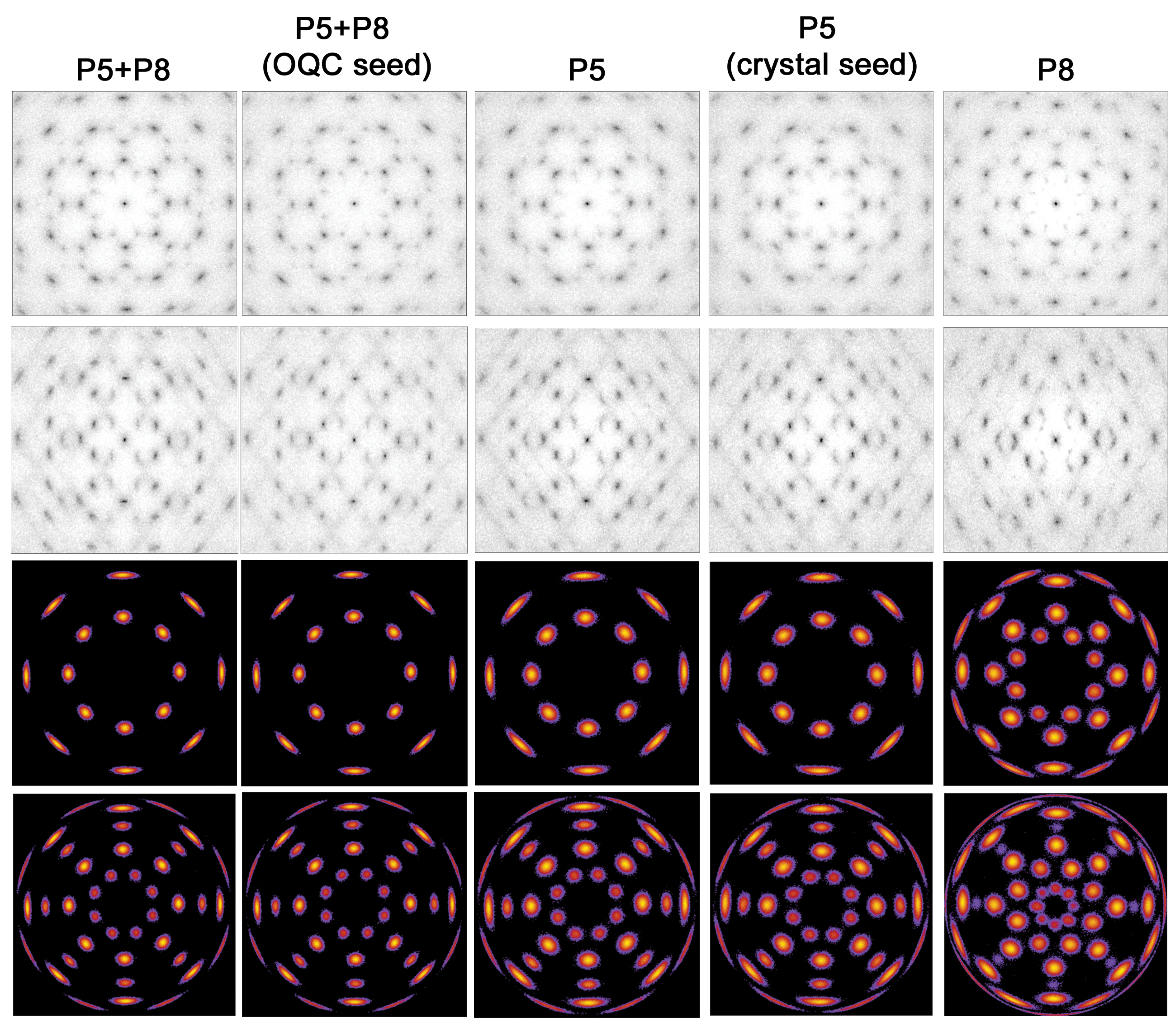}
\caption{\label{fig:compare} 
Comparison of the diffraction patterns projected onto a plane perpendicular to the 8-fold axis (top) and the 2-fold axis (bottom) for the three model designs considered. In the latter view, the periodic direction is vertical. Results for the seeded systems are also shown. BOODs calculated using an energy criterion (distance less than 1.5\,$\sigma_\mathrm{LJ}$ and energy less than $-0.2\,\epsilon_\mathrm{LJ}$, top row) and a distance criterion (distance less than 1.35\,$\sigma_\mathrm{LJ}$, bottom row) are also provided.
}
\end{figure*}

\begin{figure*}
\includegraphics[width=18cm]{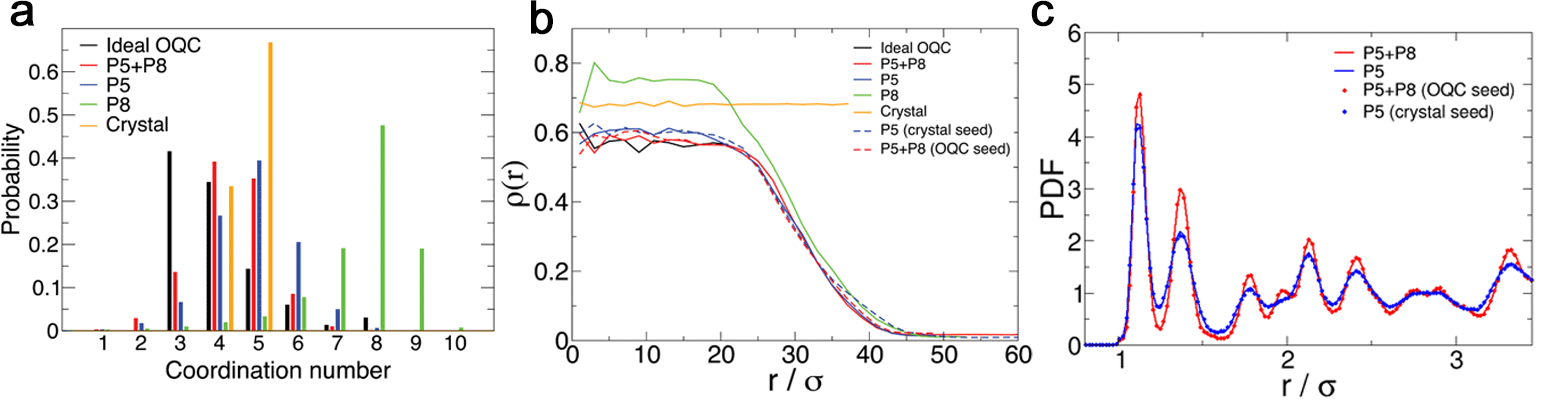}
\caption{\label{fig:additional_data} 
(a) Coordination number distributions for the assembled structures calculated using a distance criterion for bonds (distance between particles is lower than 1.30\,$\sigma_\mathrm{LJ}$). (b) Radial density of the assembled structures. For comparison, the densities of the ideal OQC and $C2/c$ crystal are also given. (c) Comparison of the radial distribution functions of the binary and P5 systems grown from the gas phase or from the OQC and $C2/c$ crystal seeds. 
}
\end{figure*}

The BOODs depicted in the main text show the geometry of the patch-patch bonds around a particle, where a distance and energy criterion were used to define a bond. It can also be useful to calculate BOODs based just on a distance criterion as this then gives a picture of the arrangement of particles (bonded and non-bonded) around a particle. These additional BOODs are shown in Fig.\ \ref{fig:compare} alongside those calculated including the energy criterion.

The geometry of the patchy particles were designed so that the next-nearest distances associated with pairs of particles at opposite corners of the ``squares'' and across the short diagonals of the ``rhombs'' in the ideal 3D octagonal QC were identical. This distance is only 1.2156 times the nearest-neighbour distance. For the assembled binary and P5 QCs, the radial distribution function does not go to zero between the first and second peaks.
Therefore, a BOOD based just on a distance criterion will show features associated with the low-distance tail of the second peak. This leads to an additional eight peaks around the equator associated with the `squares' and two sets of eight peaks at higher angles associated with the rhombs. These additional peaks explain the increase in the average coordination number (Table \ref{tab:CN}) and the coordination number distributions (Fig.\ \ref{fig:additional_data}(a)) when just a distance criterion is used.

\begin{table}
\caption{\label{tab:CN} 
Average coordination numbers, $\langle CN \rangle$, in the assembled structures and in the ideal OQC and $C2/c$ crystal.
}
\begin{ruledtabular}
\begin{tabular}{ccc}
system 
& Energy criterion
& Distance criterion
\\
\colrule
Ideal OQC & 4.0 & 4.0\\
Crystal &  4.67 & 4.67 \\
Binary  & 3.81 & 4.35 \\
Binary (OQC seed) & 3.76 & 4.36 \\
P5     & 4.03 & 4.87 \\
P5 (crystal seed)    & 4.03 & 4.90 \\
P8     & 5.93  &  7.61
\end{tabular}
\end{ruledtabular}
\end{table}

In the eight-patch system, the structure distorts from that observed in the binary and P5 systems to better make use of the extra patches. In particular, additional patchy bonds are formed to particles across the short diagonals of the ``rhombs''. This distortion involves a shortening of these distances so that they become a part of the first peak in the radial distribution function and a reorientation of the interparticle vectors so that they are closer to the angle of the patches from the vertical. These changes are directly evident in the two sets of eight peaks that appear in the BOOD compared to the binary and P5 systems. 
The peaks in the BOOD corresponding to the original bonds also move closer to the equator; the formation of the new bonds comes at the expense of the patches pointing less directly along the original bonds.
Additional consequences of these changes are that the repeat in the periodic dimension decreases (see the position of the (001) peak in the P8 diffraction patterns (Fig.\ \ref{fig:compare})), that  the puckered squares become flatter increasing the distance between diagonally opposite pairs and that the density increases (Fig.\ \ref{fig:additional_data}(b)). Hence, in the BOODs based on a distance criterion, the peaks around the equator are significantly smaller in the P8 than the binary and P5 systems, and an additional set of peaks appears in the P8 BOOD that are close to the symmetry axis due to the reduced periodic repeat.

As noted in the main text, the cluster assembled in the P8 system consists of a series of approximately crystalline domains with coherent boundaries between them that maintain the preferred orientations of the bonds. However, if one looks closely at the set of peaks associated with the new patchy bonds, one can notice that the peaks are not all of the same intensity, i.e.\ the system does not have perfect 8-fold order.
Integrating the area under the peaks confirms this conclusion and provides a more quantitative measure of the deviation from perfect symmetry (Fig.\ \ref{fig:additional_data}). Whether close to perfect 8-fold order would be restored for larger clusters, where the cluster size is much larger than the domain size of the crystallites is not clear.

We also note that the P8 crystallites have a distorted version of the $C2/c$ structure. Indeed it was the P8 simulations that led us to discover the possibility of this crystal. 

\begin{figure*}
\includegraphics[width=13cm]{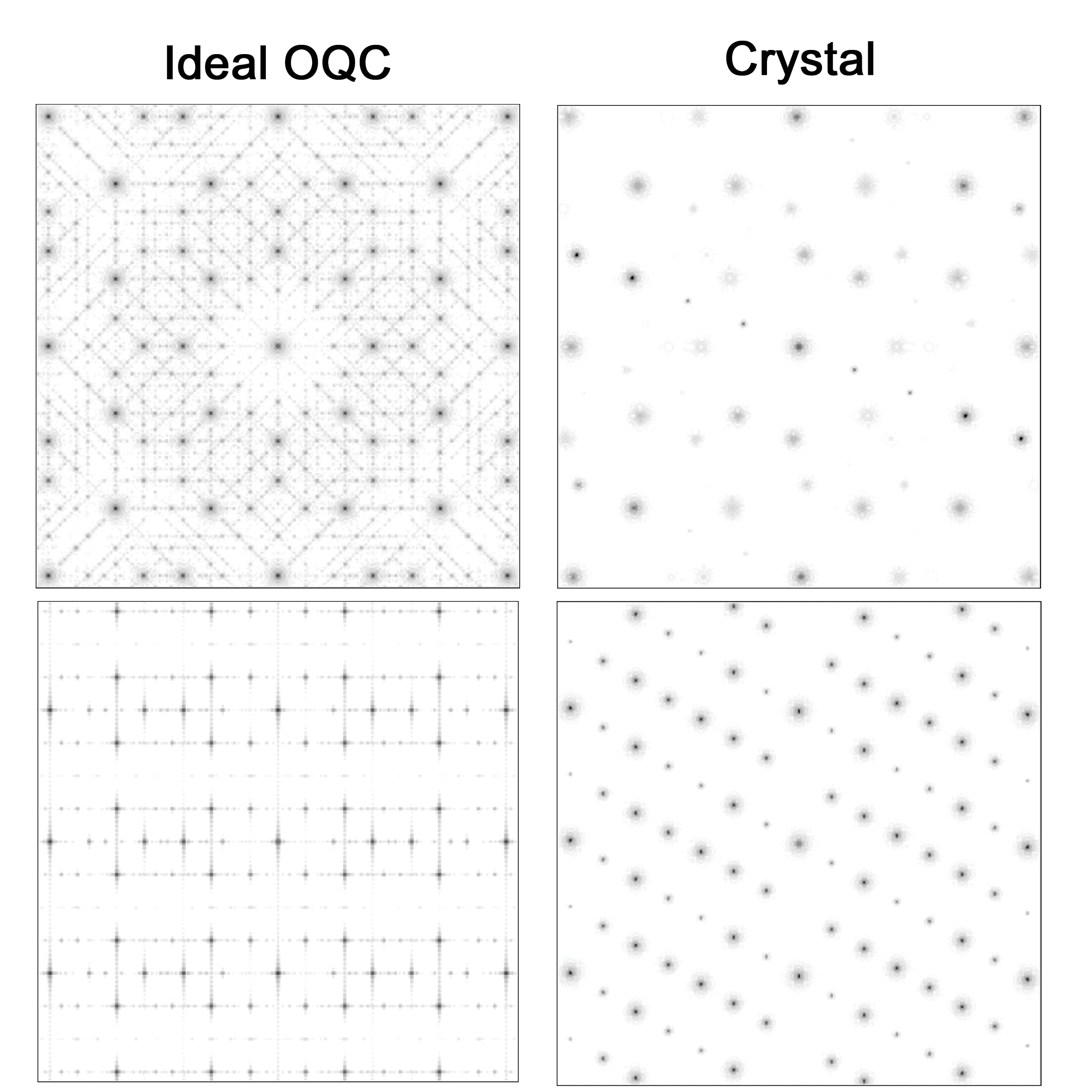}
\caption{\label{fig:patterns_ideal} 
(Left) The diffraction patterns of the ideal OQC projected onto a plane perpendicular to the 8-fold axis (top) and the 2-fold axis with the crystallographic direction aligned vertically (bottom). (Right) The diffraction patterns for the ideal $C2/c$ crystal for analogous orientations (see Fig.\ 4 in the main text).
}
\end{figure*}

The diffraction patterns for all the systems that we consider are shown in Fig.\ \ref{fig:compare} and \ref{fig:patterns_ideal}. The diffraction patterns for the binary and P5 systems are essentially identical and show clear eight-fold order and quasiperiodicity. The comparison to the ideal quasicrystal is interesting. Unsurprisingly, given that the ideal configuration is not subject to thermal noise or any type of disorder, many more peaks are observed, and the peaks observed in the assembled QCs are a subset of those for the ideal QC. However, the patterns of intensities do show differences with the ring of octagons much more apparent in the assembled systems. Presumably, these are a consequence of the differences in the local structure already highlighted in the main text. The pattern for the crystal also provides an interesting comparison. Viewed down the pseudo-eightfold axis it shows many similar peaks to the assembled QCs albeit with a clear breaking of the symmetry (the pattern only has twofold symmetry) both in the positions and intensities of those peaks.

The diffraction patterns for the P8 system are very similar to those for the binary and P5 quasicrystals (Fig.\ \ref{fig:compare}). The most discernible difference is the stretching of the pattern in the periodic direction due to the shorter repeat in that direction. Also, there are small differences in the eight-fold pattern, with some of the outer peaks in the ring of octagons less intense and the diffuse scattering near to the centre of the pattern more sharply defined.

\begin{figure*}
\includegraphics[width=14cm]{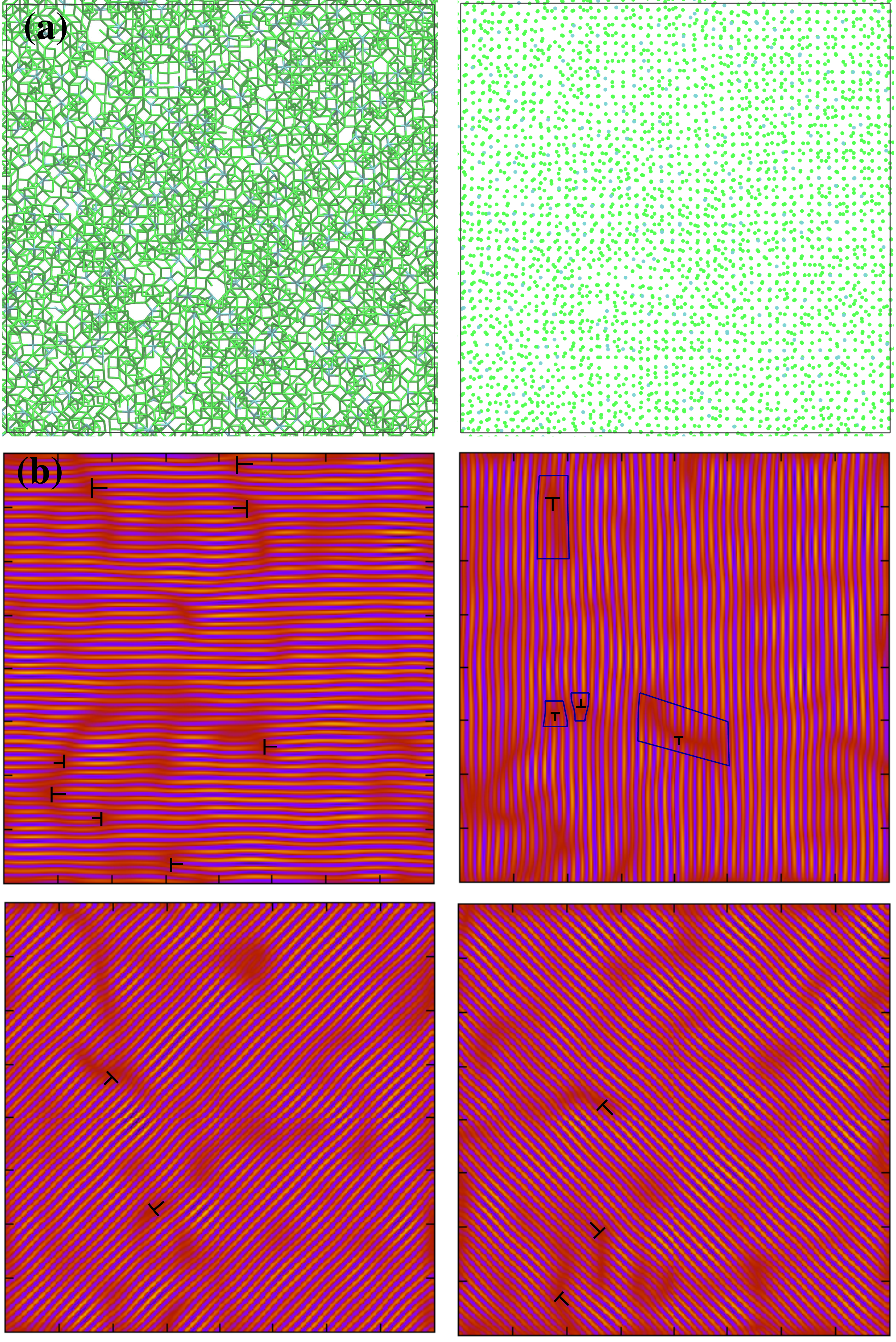}
\caption{\label{fig:dislocation} 
(a) Two representations of a slab of dimensions of $20\times 20\times 4$ (in units of $\sigma_\mathrm{LJ}$) from the binary quasicrystal. If viewed at a low angle, it becomes easier to see the lines of points in the right-hand image and to spot the dislocations.
(b) Inverse Fourier-transformed images of two equal and opposite diffraction spots in the Fourier transform of the above slab. The four images correspond to the four pairs of diffraction spots in the first intense ring and provide a representations of the lines of particles in the four equivalent directions in the octagonal quasicrystal. Isolated edge dislocations are indicated by a `T'. For the top right-hand image, circuits around the dislocations are drawn in blue. If these are followed the number of lines on one side of the dislocation will be different from the other. Features that may be interpreted in terms of pairs of nearby dislocation with equal and opposite sign are generally not highlighted.
}
\end{figure*}

\begin{figure*}[t]
\includegraphics[width=15cm]{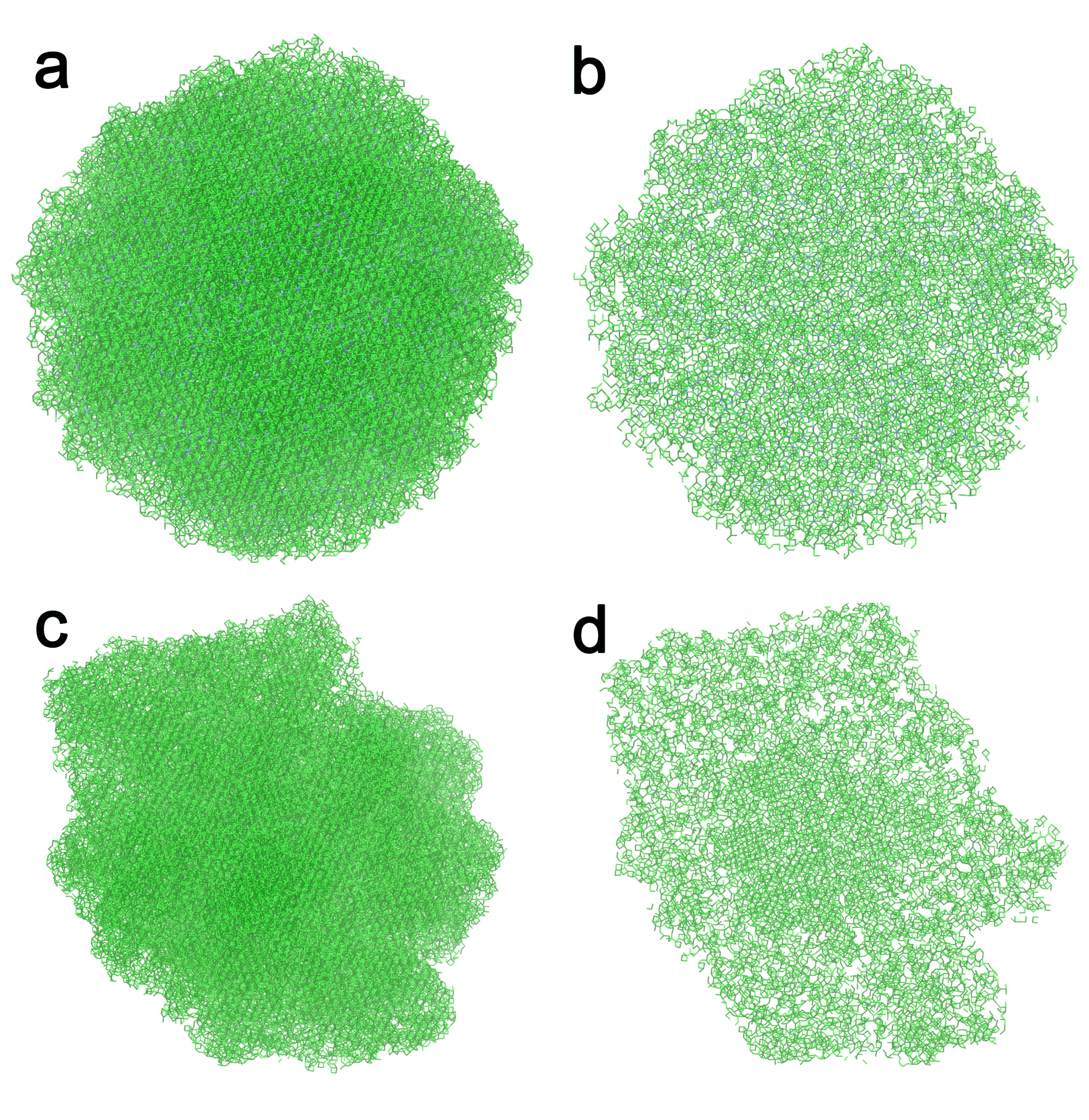}
\caption{\label{fig:snap2} Snapshots of the clusters grown from: (a) and (b) a 279-particle ideal OQC seed,  and (c) and (d) a 351-particle $C2/c$ crystal seed. Only bonds between particles (defined with a distance criterion in this figure) are shown. (a) and (c) View of the whole cluster down the 8-fold axis. (b) and (d) View of a cut of width 5\,$\sigma_\mathrm{LJ}$ of the same cluster.}
\end{figure*}

\begin{figure*}[t]
\includegraphics[width=15cm]{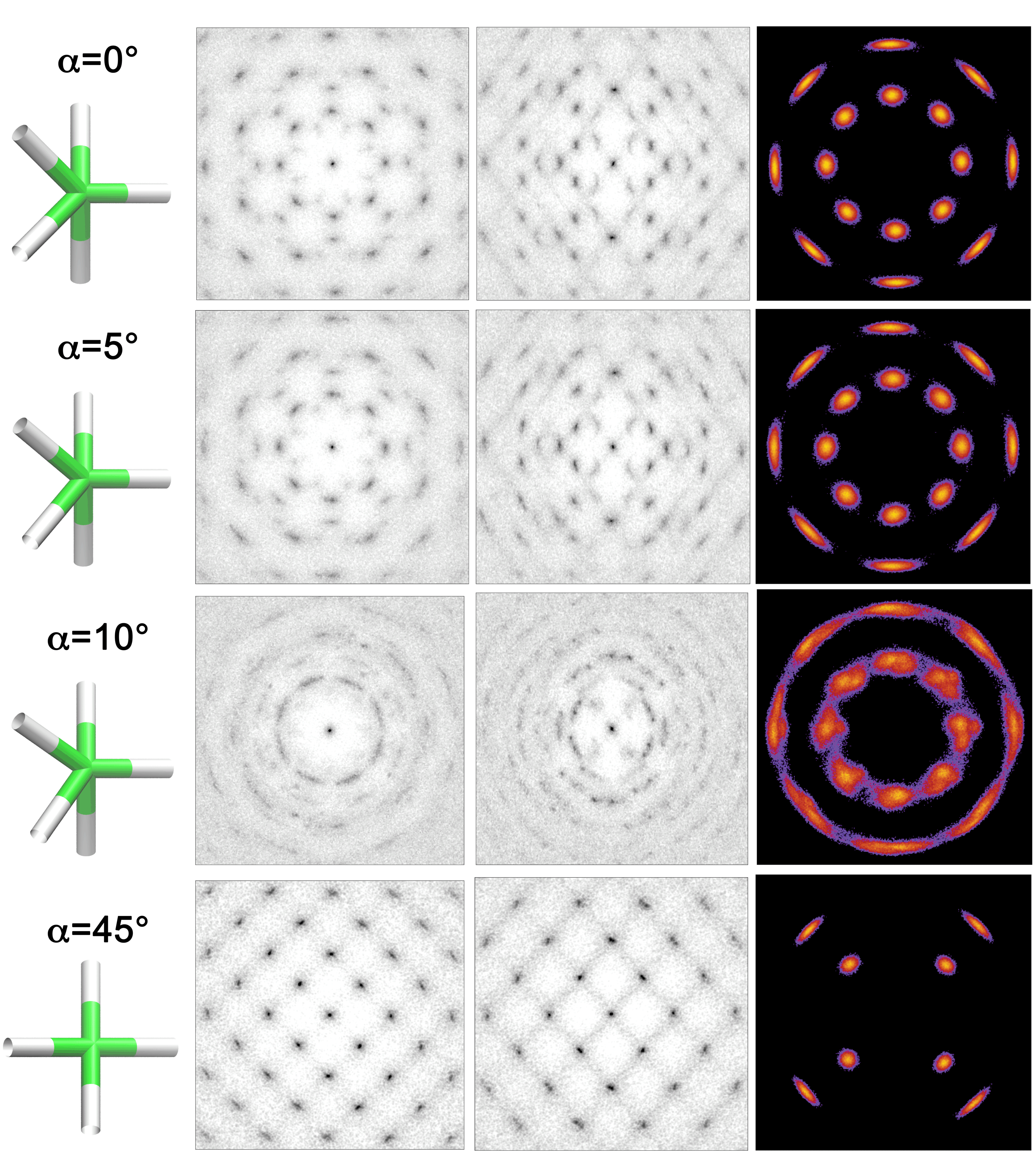}
\caption{\label{fig:summary_chiral} Effect of deviations in the patch geometry on octagonal QC assembly for the one-component P5 system. The upward- and downward facing patches are rotated by an angle $\alpha$ with respect to each other. The figure depicts the patch geometry, diffraction patterns and BOOD for four cases varying from the unperturbed $\alpha$=0 system to $\alpha=45$ where the particle is again achiral and a body-centred-cubic crystal assembles.
For $\alpha$=5$^\circ$ the chiral particles can still form an octagonal QC, but by $\alpha$=10$^\circ$ the ability to form a cluster with global orientational order is beginning to break down.}
\end{figure*}

\begin{figure*}[t]
\includegraphics[width=15cm]{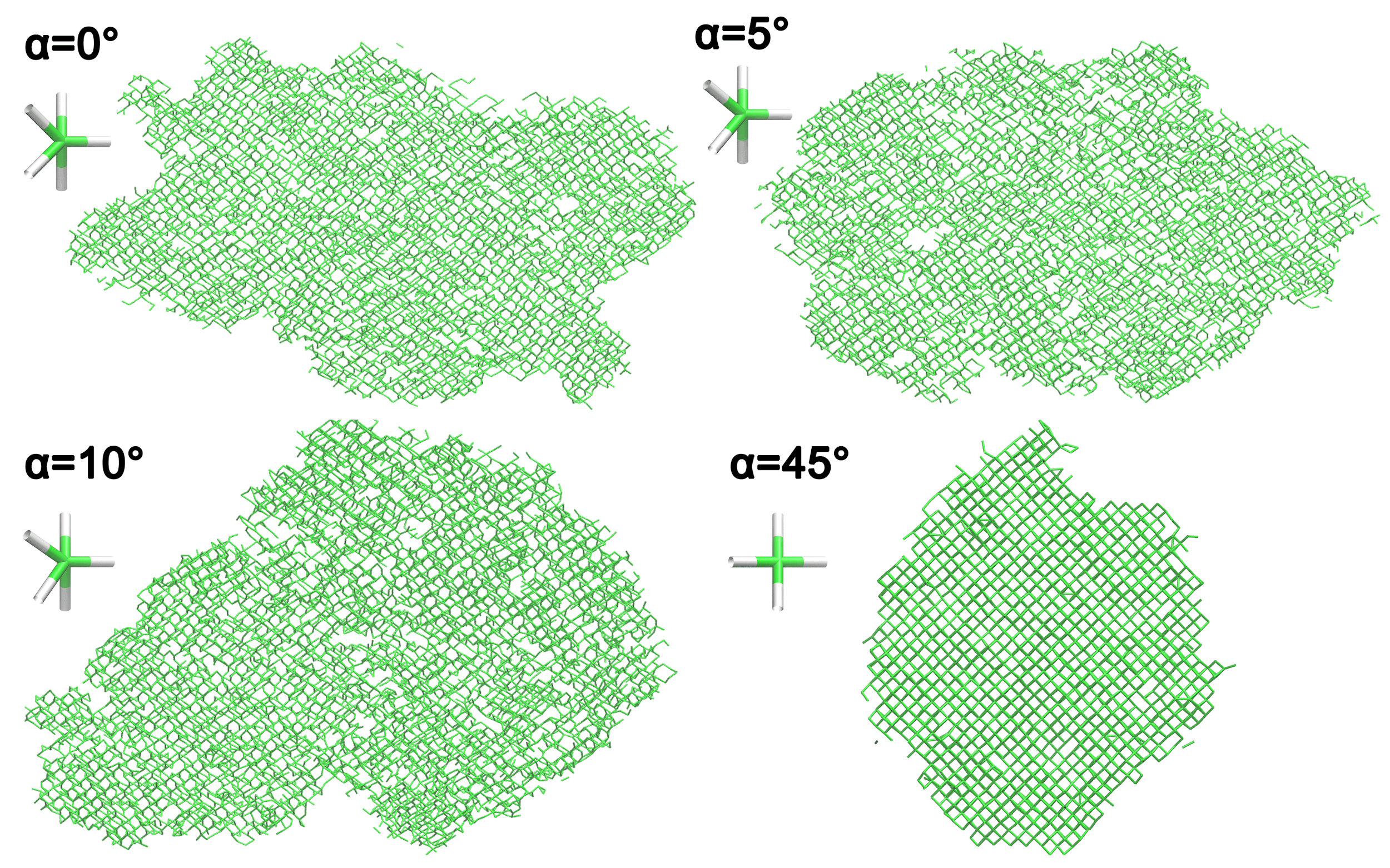}
\caption{\label{fig:slices_chiral} Slices (of width 4\,$\sigma$) though the clusters assembled for the distorted P5 systems considered in Fig.\ \ref{fig:summary_chiral}. For the first three systems the (pseudo-)eightfold axis is vertical. The $\alpha$=0 and 5$^\circ$ structures look quite similar, but for $\alpha$=10$^\circ$ there are more defects and the orientational misalignment between different ``domains'' is clear.
For $\alpha$=45$^\circ$ the system forms a body-centred-cubic crystal.
}
\end{figure*}

\section{Experimental realization}

\subsection{DNA nanotechnology}
The two most successful approaches to produce 3-dimensional periodic crystals made from DNA origami have been to use 
multi-arm DNA origami particles \cite{Posnjak24} or
DNA origami polyhedra with single-stranded ``sticky ends'' at their vertices \cite{Tian20,Liu24,Kahn25b}.
In the latter approach each edge of the polyhedron is a multi-helix bundle and the single strands that mediate the interactions have a linker section and a complementary binding region. Due to the flexible linkers, the strands mediate little further angular constraints on the vertex-vertex interactions. Thus, as torsional interactions were a prerequisite for observing octagonal QC formation in our patchy particle systems, this approach is not so well-suited to realize DNA origami analogues of our systems.

As an example of the former approach, in Ref.\ \citenum{Posnjak24} 4-arm DNA origami particles were able to assemble into a diamond lattice. Each arm consisted of a 24-helix bundle. At the centre of the design each arm splits into three 8-helix bundles that merge with the neighbouring arms. ``Insertions'' and ``deletions'' were incorporated into the arms at these points to facilitate the required bending \cite{Dietz09}. The symmetry of the design ensures the tetrahedrality of the particles. Six of the helices at the ends of the arms had short single-stranded extensions that mediated the interarm bonding; the pattern of extensions was designed to provide torsional control, in particular to ensure the ``staggered'' bonding geometry required for the diamond lattice.

This basic approach could be replicated to produce an eight-arm particle with the desired symmetry. The numbers of insertions and deletions at the centre should be tuned to ensure a sensible cone angle for the ``up'' and ``down'' arms. (Computational modelling, using for example the oxDNA model \cite{Snodin19}, could be helpful in this fine-tuning.) Although we have chosen a particular value for this angle (namely 59.3$^\circ$) in our patchy particle simulations, we do not expect octagonal QC formation to be particular sensitive to this value. Furthermore, for our patchy particles the main factor that will likely limit the range of possible values is the isotropic steric repulsions of the particles. However, the DNA origami are anisotropic in shape and have a significantly reduced excluded volume, so this constraint is likely to be less relevant in these systems. 

To make DNA origami analogues of the P8 system, one would need to functionalize the ends of the eight arms with appropriate single strands so that all arms could interact and have the appropriate torsional constraint. To make equivalents of the P5 particles, the appropriate 3-arms would need to be made non-interacting; these arms could also be made shorter to avoid potential unwanted excluded volume interactions.

An alternative approach to make an eight-arm particles might be to use DNA origami designs similar to those suggested in Ref.\ \cite{Noya21}. In particular, one could make a particle with a square-antiprism polyhedron at its core and helix-bundle extensions from each vertex. The ends of these arms could then be functionalized in a similar way to that suggested above. Although such DNA origami polyhedra are a common building block \cite{Veneziano16}, the stability of such a multi-arm design has not yet been experimentally demonstrated.

\subsection{Protein design}

Another possibility to realize the quasicrystals in this paper would be through the tools of protein design, which have advanced considerably in recent years, especially as a result of approaches that leverage the power of machine learning \cite{Winnifrith24}.

Although comparatively rare, there are still a significant number of examples of proteins that have evolved to crystallize for some functional purpose \cite{Doye06b,Schonherr18}. Similarly, proteins can be designed to form two- and three-dimensional periodic crystals \cite{Yeates17,Zhu21b}.
A recent example of the state-of-the-art in the design of proteins that can crystallize comes from Ref.\ \cite{Li23}. The approach is to first design proteins that can form high-symmetry complexes that match the local symmetry of high-symmetry sites in a crystal. Then inter-complex interactions are designed that allow these complexes to further assemble into the desired crystal. One example of this hierarchical approach was the creation of a body-centred-cubic lattice in which a binary 48-protein complex with octahedral symmetry further assembles via intercomplex interactions along the eight 3-fold axes of the octahedral complex \cite{Li23}.

One potential way to generate an analogue of the P8 particles would be to use an octameric complex with $D_4$ symmetry. Note, however, that the lower symmetry compared to the P8 particles ($D_4$ rather than $D_{4d}$) means that the rotation angle between the two square complexes that make up the octamer will not necessarily be 45$^\circ$. Instead, this angle would need to be fine-tuned in the design process to be sufficiently close to 45$^\circ$ (note, our simulations of distorted particles suggest within 5$^\circ$ might be sufficient). An additional monomer-monomer interaction would then need to be designed that facilitates the intercomplex interactions with the required directionality and preferred torsional angle.

However, it would be preferable to create analogues of the P5 particles for which we observed a much clearer propensity for octagonal QC formation. However, the lower symmetry of these particles makes this a more complex task.
One approach would be to introduce some kind of programmed symmetry breaking into the above ``P8'' octahedral complex such that only five of the eight proteins would be able to mediate inter-complex interactions. This would require the use of multiple homologous proteins with recoded intracomplex interactions to generate the necessary patterning. 
However, as the P5 particle has no rotational symmetry, eight distinct proteins would be required for the complex.
We note that such programmed symmetry breaking has been recently demonstrated for a variety of complexes \cite{Gladkov24,Lee25}. Furthermore, there has also been recent progress in generating designed multi-component complexes by the use of standardized protein building blocks \cite{Huddy24}. 

%